\documentclass[showpacs,prd,twocolumn,floatfix,nofootinbib,superscriptaddress]{revtex4-1}
\usepackage{booktabs}
\usepackage{amsmath}
\usepackage{siunitx}
\usepackage{url}
\usepackage{longtable}
\usepackage{graphicx}
\usepackage{verbatim}

\begin{document}

%-----------------------------------o
%  Front page stuff
%-----------------------------------o

\title{ $B_s \rightarrow D_s \; \ell \, \nu$ Form Factors  
and the Fragmentation Fraction Ratio $f_s/f_d$ }

\author{Christopher J.~Monahan}
\affiliation{New High Energy Theory Center and Department of Physics and 
Astronomy, \\Rutgers, the State University of New Jersey,
136 Frelinghuysen Road, Piscataway, New Jersey 08854, USA}
\affiliation{Department of Physics and Astronomy, University of Utah, 
Salt Lake City, Utah 84112, USA}
\author{Heechang Na} 
\affiliation{Ohio Supercomputer Center, 1224 Kinnear Road, Columbus, OH 
43212, USA}
\affiliation{Department of Physics and Astronomy, University of Utah, 
Salt Lake City, Utah 84112, USA}
\author{Chris M.~Bouchard}
\affiliation{School of Physics and Astronomy, University of Glasgow, Glasgow 
G12 
8QQ, UK}
\affiliation{Department of Physics and Astronomy, College of William 
and Mary, Williamsburg,
 Virginia 23187, USA}
\author{{G.~Peter} Lepage}
\affiliation{Laboratory of Elementary Particle Physics,
Cornell University, Ithaca, New York 14853, USA}
\author{Junko Shigemitsu}
\affiliation{Department of Physics,
The Ohio State University, Columbus, Ohio 43210, USA}

\collaboration{HPQCD Collaboration}
\noaffiliation
\date{\today} 

%-----------------------------------o
%  Abstract
%-----------------------------------o

\begin{abstract}
We present a lattice quantum chromodynamics determination of the scalar and vector form factors for the $B_s \rightarrow D_s \ell \nu$ decay over the full physical range of momentum transfer. In conjunction with future experimental data, our results will provide a new method to extract $|V_{cb}|$, which may elucidate the current tension between exclusive and inclusive determinations of this parameter. Combining the form factor results at non-zero recoil with recent HPQCD results for the $B \rightarrow D \ell \nu$ form factors, we determine the ratios $f^{B_s \rightarrow D_s}_0(M_\pi^2) / f^{B \rightarrow D}_0(M_K^2) = 1.000(62)$ and $f^{B_s \rightarrow D_s}_0(M_\pi^2) / f^{B \rightarrow D}_0(M_\pi^2) = 1.006(62)$. These results give the fragmentation fraction ratios $f_s/f_d = 0.310(30)_{\mathrm{stat.}}(21)_{\mathrm{syst.}}(6)_{\mathrm{theor.}}(38)_{\mathrm{latt.}} $ and $f_s/f_d = 0.307(16)_{\mathrm{stat.}}(21)_{\mathrm{syst.}}(23)_{\mathrm{theor.}}(44)_{\mathrm{latt.}}$, respectively. The fragmentation fraction ratio is an important ingredient in experimental determinations of $B_s$ meson branching fractions at hadron colliders, in particular for the rare decay ${\cal B}(B_s \rightarrow \mu^+ \mu^-)$. In addition to the form factor results, we make the first prediction of the branching fraction ratio $R(D_s) = {\cal B}(B_s\to D_s\tau\nu)/{\cal B}(B_s\to D_s\ell\nu) = 0.301(6)$, where $\ell$ is an electron or muon. Current experimental measurements of the corresponding ratio for the semileptonic decays of $B$ mesons disagree with Standard Model expectations at the level of nearly four standard deviations. Future experimental measurements of $R(D_s)$ may help understand this discrepancy.
\end{abstract}

%\pacs{12.38.Gc,
%13.20.Fc, % ($D$) 
%13.20.He } % ($B$)
%}

\maketitle

%================================= BODY =================================8

\section{Introduction}

Studies of $B$ and $B_s$ meson decays at the Large Hadron Collider provide 
precision tests of the Standard Model of particle physics and are an 
important tool in the search for new physics. For example, the first 
observation of 
the rare decay $B_s \rightarrow \mu^+ 
\mu^-$, through a combined analysis by the LHCb and CMS collaborations 
\cite{CMS:2014xfa,Aaij:2017vad}, tested the Standard Model prediction of the 
branching fraction. This decay is doubly-suppressed in the Standard Model, but 
may have large contributions from physics beyond the Standard Model (see, for 
example, 
\cite{Ge:2016lwe}). Although the observed branching fraction is currently 
consistent with Standard 
Model expectations, there is still considerable room for 
new physics, given the experimental and theoretical 
uncertainties. Both LHCb and CMS are expected to reduce their errors 
significantly in Run II and tightening
constraints on possible new physics requires a corresponding improvement in the 
theoretical determination of the Standard Model branching fraction.

Extraction of the $B_s$ meson branching fraction 
${\cal B}( B_s \rightarrow \mu^+ \mu^-)$ relies on the normalization 
channels $B^+_u \rightarrow J/\Psi(\mu^+\mu^-) K^+$ and 
$B^0_d \rightarrow K^+ \pi_-$ \cite{Adeva:2009ny}.
The branching fraction can then be expressed as \cite{CMS:2014xfa}
\begin{equation}
{\cal B}(B_s \rightarrow \mu^+ \mu^-) = {\cal B}(B_q \rightarrow X) 
\frac{f_q}{f_s} \frac{\epsilon_X}{\epsilon_{\mu \mu}} \frac{N_{\mu \mu}}
{N_X},
\end{equation}
where the $f_q$ are the fragmentation fractions, which give the 
probability that a $b$-quark hadronizes into a $B_q$ meson. The $\epsilon$ 
factors in this equation represent detector efficiencies and the $N$ 
factors denote the observed numbers of events. 

The analysis of \cite{CMS:2014xfa} used the value of $f_s/f_d = 
0.259(15)$, determined from LHCb experimental data 
\cite{Aaij:2011jp,Aaij:2013qqa,LHCb:2013lka}.
The ratio $f_s/f_d$
depends on the kinematic range of the experiment, leading to the introduction 
of an additional systematic uncertainty in the value of $f_s/f_d$ to account
for the extrapolation of the LHCb result to the CMS acceptance. Reducing 
sources of systematic uncertainties in the value of this ratio will 
improve the precision of the determination of the $B_s \rightarrow \mu^+ \mu^-$ 
branching fraction. Indeed, an accurate value for the fragmentation fraction 
ratio is necessary for improved measurements of other $B_s$ 
meson decay branching fractions at the LHC \cite{Adeva:2009ny}. 

The ratio of the fragmentation fractions, $f_s/f_d$, can be expressed in terms 
of the ratios of form factors \cite{Fleischer:2010ay,Fleischer:2010ca},
\begin{equation}\label{eq:ffrat}
{\cal N}_F = \left [ \frac{f_0^{(s)}(M_\pi^2)}{f_0^{(d)}(M_K^2)} \right ]^2 
\quad \mathrm{and} \quad 
{\cal N}'_F = \left [ \frac{f_0^{(s)}(M_\pi^2)}{f_0^{(d)}(M_\pi^2)}
 \right ]^2,
\end{equation}
where $f_0^{(q)}(M^2)$  is the scalar form factor of the 
$B_q \rightarrow D_q l \nu$ semileptonic decay at $q^2 = M^2$. The first 
lattice 
calculations of the form factor ratios in Equation \eqref{eq:ffrat} using heavy 
clover bottom and charm quarks were published in \cite{Bailey:2012rr}. In 
addition, the form 
factors, $f_+(q^2)$ and $f_0(q^2)$, for 
the semileptonic decay $B_s \rightarrow D_s \ell \nu$ were determined with 
twisted mass fermions for the region near zero recoil in \cite{Atoui:2013zza}.

In this article we calculate the form factors, $f_+(q^2)$ and $f_0(q^2)$, for 
the semileptonic decay $B_s \rightarrow D_s \ell \nu$. We present a
determination of these form factors over the full physical range of momentum 
transfer, $q^2$ using the modified $z$-expansion for the chiral-continuum-kinematic extrapolation. We combine these form factor results with recent HPQCD results 
for the $B \rightarrow D \ell \nu$ decay \cite{Na:2015kha} to
determine the ratios of $B_s \rightarrow D_s \ell \nu$ and $B 
\rightarrow D \ell \nu$ form factors relevant to the ratio of fragmentation 
fractions, $f_s/f_d$.  

We use the non-relativistic (NRQCD) action for the bottom quarks and the Highly 
Improved Staggered Quark (HISQ) action for the charm quarks. 
Our form factors for $B \rightarrow D \ell \nu$ have appeared already 
in \cite{Na:2015kha}.  
Here we first present $B_s \rightarrow D_s \ell \nu$ form 
factor results and then proceed to the form factor ratios.
We find
\begin{equation}
\label{eq:ffratresults}
\frac{f_0^{(s)}(M_\pi^2)}{f_0^{(d)}(M_K^2)} = 1.000(62)
\quad {\rm and} \quad 
\frac{f_0^{(s)}(M_\pi^2)}{f_0^{(d)}(M_\pi^2)} = 1.006(62).
\end{equation}
This leads to
\begin{equation}
\frac{f_s}{f_d} = 
0.310(30)_{\mathrm{stat.}}(21)_{\mathrm{syst.}}(6)_{\mathrm{theor.}}(38)_{
\mathrm { latt. } }
\end{equation}
and
\begin{equation}\label{eq:fsfdpp2}
\frac{f_s}{f_d} = 
0.307(16)_{\mathrm{stat.}}(21)_{\mathrm{syst.}}(23)_{\mathrm{theor.}}(44)_{
\mathrm{latt.}},
\end{equation}
respectively. The uncertainties in these results are: the experimental 
statistical and systematic uncertainties; theoretical uncertainties 
(predominantly arising from a factor that captures deviations from naive 
factorization and, in Equation \eqref{eq:fsfdpp2}, an electroweak correction 
factor); and the uncertainties in 
our lattice input. In quoting these results, we have 
assumed that there are no correlations between the lattice 
results and the other sources of uncertainty.

In addition to determining the fragmentation fraction ratio relevant to the 
measurement of the branching fraction for the rare decay, $B_s \rightarrow 
\mu^+ \mu^-$, the semileptonic $B_s\to D_s\ell\nu$ decay provides a new method 
to determine the CKM matrix element $|V_{cb}|$. There is a 
long-standing tension between determinations of $|V_{cb}|$ from exclusive 
and inclusive measurements of the 
semileptonic $B$ meson decays (see, for example, 
\cite{Amhis:2014hma,hfag:2016smi} and the 
review in \cite{pdg15}), although recent analyses suggest the tension has
eased \cite{Gambino:2016jkc,Bigi:2016mdz}. The $B_s\to D_s\ell\nu$ decay has 
yet to be observed experimentally and consequently has received less 
theoretical attention than semileptonic decays of the $B$ meson. The studies 
that have been undertaken for the $B_s\to D_s\ell\nu$ decay
include calculations based on relativistic quark models 
\cite{Zhao:2006at,Bhol:2014jta}, light-cone sum rules \cite{Li:2009wq}, 
perturbative factorization \cite{Fan:2013kqa} and estimates using the 
Bethe-Salpeter method \cite{Li:2010bb,Chen:2011ut}. At present, there is one 
unquenched lattice calculation of the form factor ${\cal G}(1)$ at zero recoil 
\cite{Atoui:2013zza}. The FNAL/MILC collaboration has previously studied the ratio of 
the form factors of the $B_s\to D_s\ell\nu$ and $B\to D\ell\nu$ decays 
\cite{Bailey:2012rr}.

We determine the form factor for the $B_s\to D_s\ell\nu$ semileptonic decay at 
zero momentum transfer to be $f_0(0) = f_+(0) = 0.656(31)$ and at zero 
recoil 
to be ${\cal G}(1)\propto f_+(q^2_{\mathrm{max}}) = 1.068(40)$. Although 
experimental data is frequently presented in the form $|V_{cb}|{\cal G}(1)$, 
the 
additional information provided by our calculation of the shape of the form 
factors throughout the kinematic range will, when combined with future 
experimental data, provide a new 
method to 
extract $|V_{cb}|$ and may elucidate the puzzle of the tension 
between inclusive and exclusive determinations of this CKM matrix element.

In the next section we briefly outline the details of the calculation, 
including the gauge ensembles, bottom-charm currents and two- and 
three-point correlator construction. Our calculation closely parallels that 
presented in \cite{Na:2015kha} for the $B \rightarrow D\ell \nu$ semileptonic 
decay and we refer the reader to that work for further details. In Section 
\ref{sec:corrfits} we discuss correlator fits to our lattice data and Section 
\ref{sec:zexp} covers the chiral-continuum-kinematic extrapolations, which 
follows closely the methodology of \cite{Na:2015kha}. We explain how some of
the correlations between the new $B_s \rightarrow D_s \ell \nu$ data and 
the $B \rightarrow D\ell \nu$ data are incorporated into the 
chiral-continuum-kinematic expansion. 
Section \ref{sec:results} presents our final results for the $B_s \rightarrow 
D_s \ell \nu$ form factors, for ${\cal N}_F$ and $\tilde{{\cal N}}_F$, 
and for $f_s/f_d$ and $R(D_s)$. We summarize in Section \ref{sec:summary} and 
in Appendix 
\ref{sec:ffdetails} we give the information necessary to reconstruct the $B_s 
\rightarrow D_s \ell \nu$ 
form factors. The analogous details for $B \rightarrow D \ell \nu$ form factors 
were summarized in Appendix A of \cite{Na:2015kha}.

\section{\label{sec:lattsetup}Ensembles, currents and correlators}

Our determination of the form factors for the $B_s \rightarrow D_s \ell \nu$ 
semileptonic decay closely parallels the analysis presented in 
\cite{Na:2015kha}. Here we simply sketch the key ingredients of the analysis 
and refer the reader to Sections II and III of \cite{Na:2015kha} for more 
details of the lattice calculation.

We use five gauge ensembles, summarized in Table \ref{tab:milc}, generated by 
the MILC collaboration \cite{Bazavov:2009bb}. These ensembles include three 
``coarse'' (with 
lattice spacing $a \approx \SI{0.12}{fm}$) and two ``fine'' (with $a \approx 
\SI{0.09}{fm}$) ensembles and incorporate $n_f = 2+1$ flavors of AsqTad sea 
quarks.
\begin{table}
\caption{\label{tab:milc}
Simulation details on three ``coarse'' and two ``fine''  $n_f = 2 + 1$ MILC 
ensembles.
}
\begin{ruledtabular}
\begin{tabular}{cccccc}
Set &  $r_1/a$ & $m_l/m_s$ (sea)   &  $N_{\mathrm{conf}}$&
$N_{\mathrm{tsrc}}$ & $L^3 \times N_t$ \\
\vspace*{-10pt}\\
\hline 
\vspace*{-10pt}\\
C1  & 2.647 & 0.005/0.050   & 2096  &  4 & $24^3 \times 64$ \\
C2  & 2.618 & 0.010/0.050  & 2256   & 2 & $20^3 \times 64$ \\
C3  & 2.644 & 0.020/0.050  & 1200  & 2 & $20^3 \times 64$ \\
F1  & 3.699 & 0.0062/0.031  & 1896  & 4  & $28^3 \times 96$ \\
F2  & 3.712 & 0.0124/0.031  & 1200  & 4 & $28^3 \times 96$ \\
\end{tabular}
\end{ruledtabular}
\end{table}
In addition, we tabulate the light pseudoscalar masses on these ensembles, for 
both AsqTad and HISQ valence quarks, in Table \ref{tab:deltapi}. The 
difference in these masses captures discretization effects arising from partial 
quenching. We account for these effects in the 
chiral-continuum-kinematic expansion, which we discuss in more detail in 
Section 
\ref{sec:zexp}.
\begin{table*}
\caption{\label{tab:deltapi}
Meson masses on MILC ensembles for both AsqTad \cite{Bazavov:2009bb} and HISQ 
valence quarks \cite{Bouchard:2014ypa}. The $aM_{\eta_s}$ values are determined 
with 
HISQ valence quarks in \cite{Bouchard:2014ypa}.}
\begin{ruledtabular}
\begin{tabular}{cccccc}
Set & $M_\pi^{\mathrm{AsqTad}}$   & $aM_\pi^{\mathrm{HISQ}}$ & 
$aM_K^{\mathrm{AsqTad}}$ & $aM_K^{\mathrm{HISQ}}$ & $aM_{\eta_s}$\\
\vspace*{-10pt}\\
\hline 
\vspace*{-8pt}\\
C1 & 0.15971(20) & 0.15990(20) & 0.36530(29) & 0.31217(20) & 0.41111(12) \\
C2 & 0.22447(17) & 0.21110(20) & 0.38331(24) & 0.32851(48) & 0.41445(17) \\
C3 & 0.31125(16) & 0.29310(20) & 0.40984(21) & 0.35720(22) & 0.41180(23) \\
F1 & 0.14789(18) & 0.13460(10) & 0.25318(19) & 0.22855(17) & 0.294109(93) \\
F2 & 0.20635(18) & 0.18730(10) & 0.27217(21) & 0.24596(14) & 0.29315(12) \\
\end{tabular}
\end{ruledtabular}
\end{table*}

In Table \ref{tab:valq} we list the valence quark masses for the NRQCD 
bottom quarks and HISQ charm quarks \cite{Na:2012kp,Bouchard:2014ypa}. For 
completeness and ease of reference, we include 
both the tree-level wave function renormalization for the massive HISQ quarks 
\cite{Monahan:2012dq} and the spin-averaged $\Upsilon$ mass, corrected for 
electroweak effects, determined in \cite{Na:2012kp}.
\begin{table}
\caption{\label{tab:valq}
Valence quark masses $a m_b$ for NRQCD bottom quarks and 
 $a m_s$ and $a m_c$ for HISQ  strange
 and charm quarks.  The fifth
column gives 
$Z_2^{(0)}(a m_c)$, the tree-level wave function renormalization 
constant for massive (charm) HISQ quarks. The sixth column lists the values of 
the spin-averaged $\Upsilon$ mass, corrected for electroweak effects.}
\begin{ruledtabular}
\begin{tabular}{cccccc}
Set & $a m_b$   & $a m_s$ & $a m_c$ & $  Z_2^{(0)}(a m_c) $ & 
$aE_{b\overline{b}}^{\mathrm{sim}}$  \\
\vspace*{-10pt}\\
\hline 
\vspace*{-10pt}\\
C1 & 2.650 & 0.0489 & 0.6207 & 1.00495618 & 0.28356(15) \\
C2 & 2.688 & 0.0492 & 0.6300 & 1.00524023 & 0.28323(18) \\
C3 & 2.650 & 0.0491 & 0.6235 & 1.00504054 & 0.27897(20) \\
F1 & 1.832 & 0.0337 & 0.4130 & 1.00103879 & 0.25653(14) \\
F2 & 1.826 & 0.0336 & 0.4120 & 1.00102902 & 0.25558(28) \\
\end{tabular}
\end{ruledtabular}
\end{table}

To study $B_s \rightarrow D_s$ semileptonic decays, we evaluate the 
matrix element of the bottom-charm vector current, $V^\mu$, between $B_s$ and 
$D_s$ states. We express this matrix element in terms of the form factors
$f_+(q^2)$ and $f_0(q^2)$ as
\begin{align}
\langle D_s(p_{D_s}) {} & | V^\mu | B_s(p_{B_s}) \rangle = f_0(q^2) 
\frac{M_{B_s}^2-M_{D_s}^2}{q^2}q^\mu \nonumber\\
{} & + f_+(q^2) \left[
p_{B_s}^\mu + p_{D_s}^\mu - \frac{M_{B_s}^2-M_{D_s}^2}{q^2}q^\mu
\right],
\end{align}
where the momentum transfer is $q^\mu = p_{B_s}^\mu - p_{D_s}^\mu$. In practice 
it is simpler to work with the form factors $f_\parallel$ and 
$f_\perp$, which are related to $f_+(q^2)$ and $f_0(q^2)$ via
\begin{align}
f_+^{(s)}(q^2) = {} & \frac{1}{\sqrt{2M_{B_{(s)}}}}\Big[f_\parallel^{(s)}(q^2) 
\nonumber\\
{} & \qquad
+ 
(M_{B_{(s)}}-E_{D_{(s)}})f_\perp^{(s)}(q^2)\Big], \\
f_0^{(s)}(q^2) = {} & 
\frac{\sqrt{2M_{B_{(s)}}}}{M_{B_{(s)}}^2-M_{D_{(s)}}^2}\bigg[(M_{B_{(s)}}-E_{D_{
(s)} }
)f_\parallel^{(s)}(q^2) \nonumber\\
{} & \qquad + 
(E_{D_{(s)}}^2-M_{D_{(s)}}^2)f_\perp^{(s)}(q^2)\bigg].
\end{align}
Here $E_{D_s}$ is the energy of the daughter $D_s$ meson in the rest frame of 
the $B_s$ meson. In the following, we work in the rest frame of the $B_s$ meson 
and when we refer to the spatial momentum, $\vec{p}$, we mean the momentum of 
the $D_s$ meson.

NRQCD is an effective theory for heavy quarks and results determined using 
lattice
NRQCD must be matched to full QCD to make contact with experimental data. We 
match the bottom-charm currents, $J_\mu$, at one loop in perturbation theory 
through ${\cal O}(\alpha_s, \Lambda_{\mathrm{QCD}}/m_b, \alpha_s/am_b)$, where 
$am_b$ is the bare lattice mass \cite{Monahan:2012dq}. We re-scale all currents 
by the nontrivial massive wave function renormalization for the HISQ charm 
quarks, tabulated in Table \ref{tab:valq}, \cite{Na:2015kha}.

\begin{figure}
\centering
\caption{\label{fig:3pt}Lattice setup for the three-point correlators. See 
accompanying text for details.}
\includegraphics[width=0.45\textwidth,keepaspectratio]{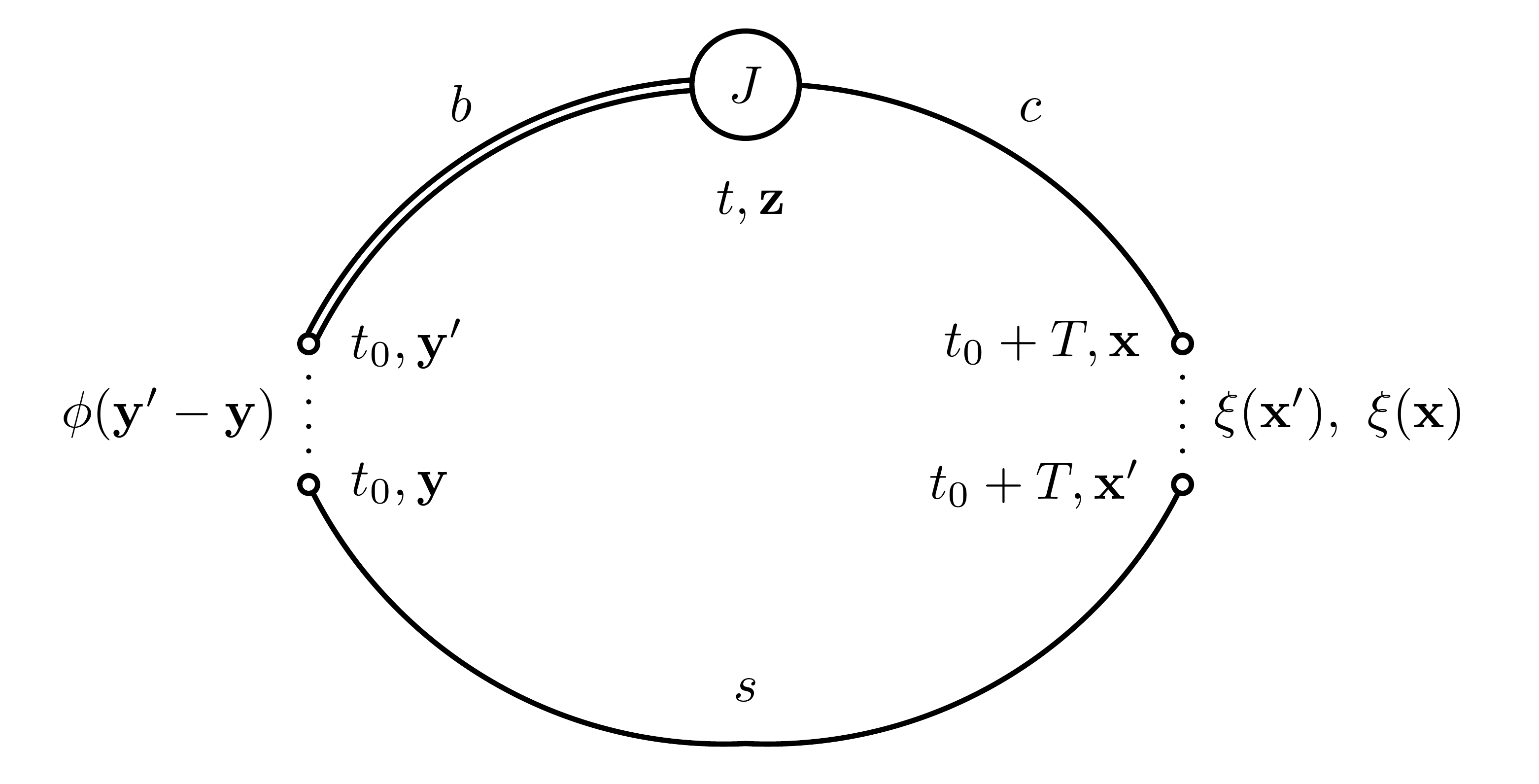}
\end{figure}

We calculate $B_s$ and $D_s$ meson two-point correlators and three-point 
correlators of the bottom-charm currents, $J_\mu$. We use smeared heavy-strange 
bilinears to represent the $B_s$ meson and incorporate both delta-function and 
Gaussian smearing, with a smearing radius of $r_0/a = 5$ and $r_0/a = 7$ on the 
coarse and fine ensembles, respectively. Three-point correlators are computed 
with the setup illustrated in Figure \ref{fig:3pt}. The $B_s$ meson is created 
at time $t_0$ and a current $J_\mu$ inserted at timeslice $t$, between $t_0$ 
and $t_0+T$. The daughter $D_s$ meson is then annihilated at timeslice 
$t_0+T$. We use four values of $T$: 12, 13, 14, and 15 on the coarse lattices; 
and 21, 22, 23, and 24 on the fine lattices. We implement spatial sums at the 
source through the $U(1)$ random wall sources $\xi(x)$ and $\xi(x')$ 
\cite{Na:2010uf}. We generate data for four different values of the $D_s$ meson 
momenta, $\vec{p} = 2\pi/(aL)(0,0,0)$, $\vec{p} = 
2\pi/(aL)(1,0,0)$, $\vec{p} = 2\pi/(aL)(1,1,0)$, and $\vec{p} = 
2\pi/(aL)(1,1,1)$, where $L$ is the spatial lattice extent.

We fit $B_s$ meson two-point functions to a sum of decaying exponentials in 
Euclidean time, $t$,
\begin{align}
C_{B_s}^{\beta,\alpha}(t) = {} & \sum_{i=0}^{N_{B_s}-1} b_i^\beta  
b_i^{\alpha\ast} e^{-E_i^{B_s,\mathrm{sim}}t} \nonumber\\
{} & + \sum_{i=0}^{N_{B_s}'-1}b_i^{\prime\,\beta}
b_i^{\prime\,\alpha\ast} 
(-1)^te^{-E_i^{\prime\,B_s,\mathrm{sim}}t}.
\end{align}
Here the superscripts $\alpha$ and $\beta$ indicate the smearing associated 
with the $B_s$ meson source (delta function or Gaussian); the $b_i$ and $b_i'$ 
are amplitudes associated with the ordinary non-oscillatory states and the
oscillatory states 
that arise in the staggered quark formalism; the meson energies are  
$E_i^{B_s,\mathrm{sim}}$ and $E_i^{\prime\,B_s,\mathrm{sim}}$ for the 
non-oscillatory and oscillatory states, respectively; and $N_{B_s}^{(\prime)}$ 
is the 
number of exponentials included in the fit.

The ground state $B_s$ energy in NRQCD, $E_0^{B_s,\mathrm{sim}}$, is related to 
the true energy in full QCD, $E_0^{B_s}$, by
\begin{equation}
E_0^{B_s} \equiv M_{B_s} = 
\frac{1}{2}\left[\overline{M}_{b\overline{b}}^{\mathrm{exp}} - 
E_{b\overline{b}}^{\mathrm{sim}}\right] +  E_0^{B_s,\mathrm{sim}},
\end{equation}
because the $b$-quark rest mass has been integrated out in NRQCD. Here 
$\overline{M}_{b\overline{b}}^{\mathrm{exp}}$ is the spin-averaged $\Upsilon$ 
mass used to tune the $b$-quark mass and $aE_{b\overline{b}}^{\mathrm{sim}}$ 
was 
determined in \cite{Na:2012kp}. We tabulate the values for 
$aE_{b\overline{b}}^{\mathrm{sim}}$ in Table \ref{tab:valq}.

We fit the $D_s$ meson two-point functions to the form
\begin{align}
C_{D_s}(t;\vec{p}) = {} & \sum_{i=0}^{N_{D_s}-1} 
|d_i|^2 \left[e^{-E_i^{D_s}t}+e^{-E_i^{D_s}(N_t-t)}\right] \nonumber\\
 +\sum_{i=0}^{N_{D_s}'-1}{} & |d_i'|^2 
(-1)^t\left[e^{-E_i^{\prime\,D_s}t}+e^{-E_i^{\prime\,D_s}(N_t-t)}\right].
\end{align}
For the three-point correlator we use the fit ansatz
\begin{align}
C_{J}^\alpha{} & (t,T;\vec{p}) = \sum_{i=0}^{N_{D_s}-1} \sum_{j=0}^{N_{B_s}-1} 
A_{ij}^\alpha e^{-E_i^{D_s}t}e^{-E_j^{B_s,\mathrm{sim}}(T-t)} \nonumber\\
{} & + \sum_{i=0}^{N_{D_s}'-1} \sum_{j=0}^{N_{B_s}-1} 
B_{ij}^\alpha 
(-1)^te^{-E_i^{\prime\,D_s}t}e^{-E_j^{B_s,\mathrm{sim}}(T-t)} 
\nonumber\\
{} & + \sum_{i=0}^{N_{D_s}-1} \sum_{j=0}^{N_{B_s}'-1} 
C_{ij}^\alpha 
(-1)^te^{-E_i^{D_s}t}e^{-E_j^{\prime\,B_s,\mathrm{sim}}(T-t)} 
\nonumber\\
{} & + \sum_{i=0}^{N_{D_s}'-1} \sum_{j=0}^{N_{B_s}'-1} 
D_{ij}^\alpha 
(-1)^{T}e^{-E_i^{\prime\,D_s}t}e^{-E_i^{\prime\,B_s,\mathrm{sim}}(T-t)} .
\end{align}
The amplitudes $A_{ij}^\alpha$ for energy levels $(i,j)$ depend on the current 
$J_\mu$, the daughter $D_s$ meson momentum $\vec{p}$, and the smearing of the 
$B_s$ meson source, $\alpha$.

The hadronic matrix element between $B_s$ and $D_s$ meson states is then given 
in terms of the ground state energies and amplitudes extracted from two- and 
three-point correlator fits by the relation
\begin{equation}
\langle D_s(\vec{p}) | V^\mu | B_s \rangle = 
\frac{A_{00}^\alpha}{d_0b_0^{\alpha\ast}}\sqrt{2a^3E_0^{D_s}}\sqrt{2a^3M_{B_s}}.
\end{equation}
For more details on this relation, see Section III of \cite{Na:2015kha}.

\section{\label{sec:corrfits}Correlator fit and form factor results}

We employ a Bayesian multi-exponential fitting procedure, based on the 
\verb+python+ packages \verb+lsqfit+ \cite{lsqfit} and \verb+corrfitter+ 
\cite{corrfitter}, that has been 
used by the HPQCD collaboration for a wide range of lattice calculations. 
Statistical correlations between data points, and correlations between data and 
priors, are automatically captured with the \verb+gvar+ class \cite{gvar}, 
which 
facilitates the straightforward manipulation of Gaussian-distributed random 
variables.

\begin{figure}
\centering
\caption{\label{fig:ds2pt}Fit results for the $D_s$ meson two-point correlator 
as a function of the number of exponentials included in the fit on ensemble F1. 
The upper plot includes data for all four values of the spatial momentum of the 
$D_s$ meson. The lower plot compares the values for the ground state energy 
from the simultaneous fit with two alternative fitting strategies, 
which are described in the text, at zero spatial momentum. Note the magnified 
scale on the vertical axis in the lower panel.}
\includegraphics[width=0.4\textwidth,keepaspectratio]{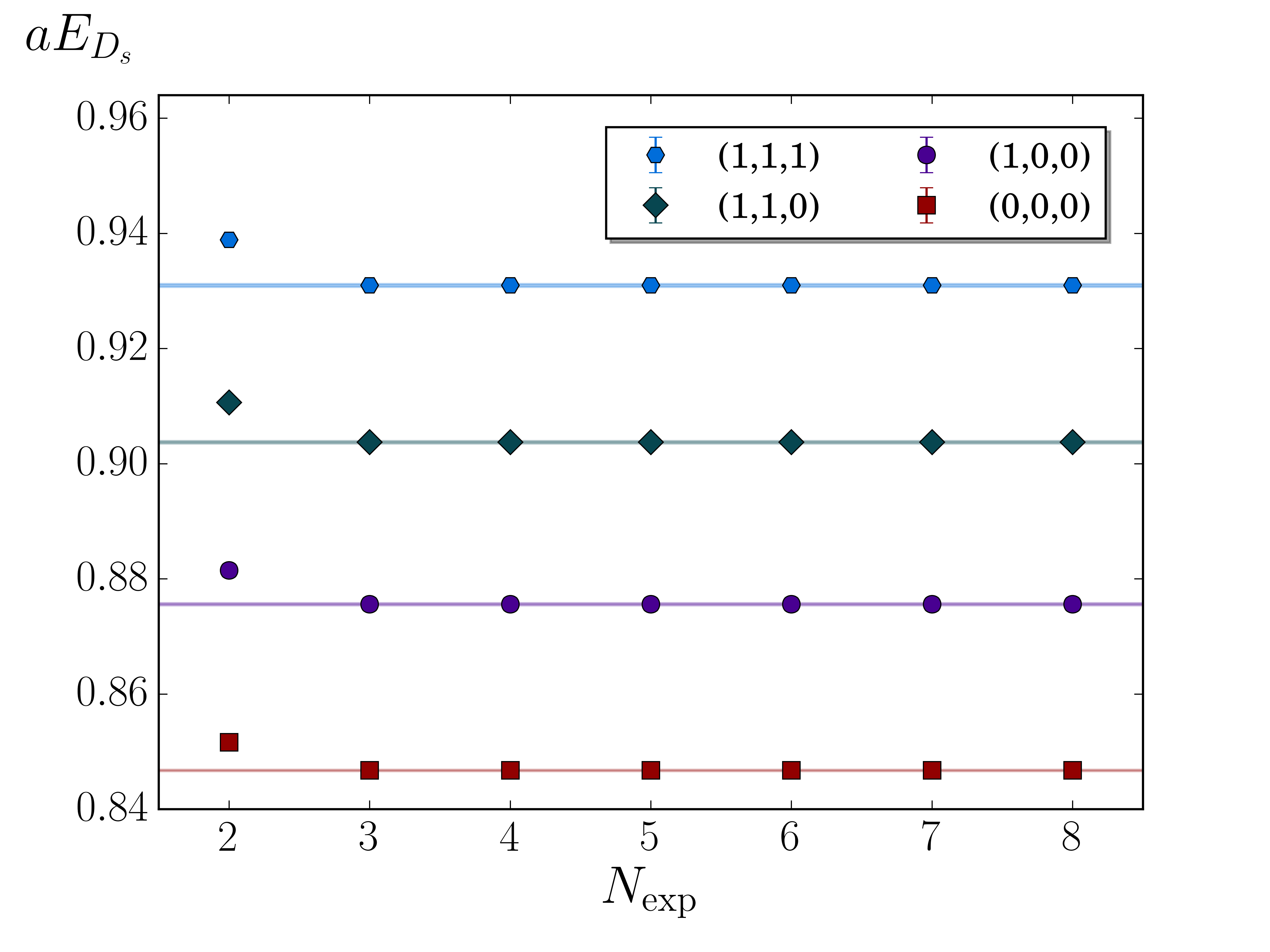}
\includegraphics[width=0.4\textwidth,keepaspectratio]{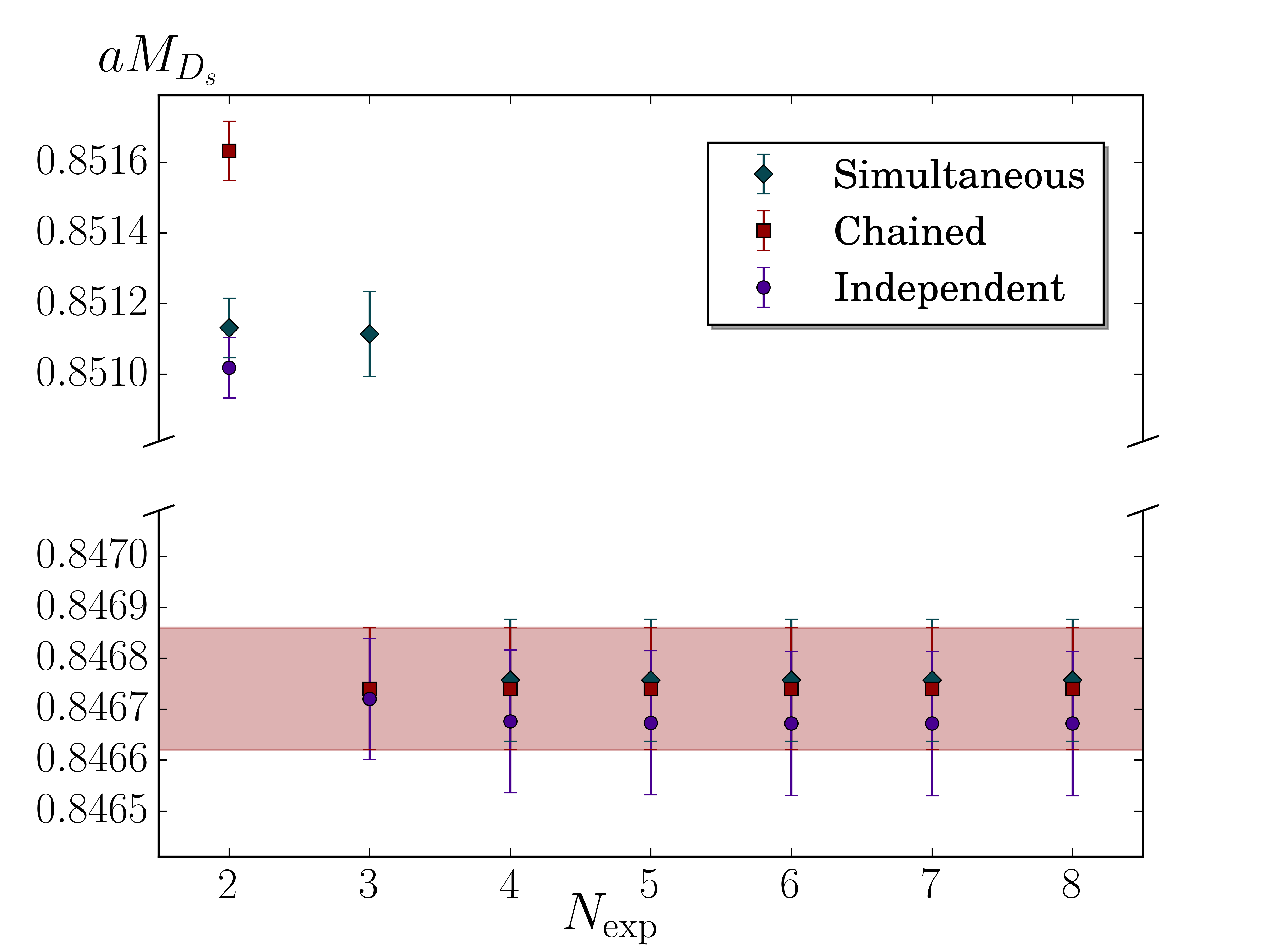}
\end{figure}

In this Bayesian multi-exponential approach, one uses a number of indicators of 
fit stability, consistency, and goodness-of-fit to check the fit 
results. For example, we check that, beyond a minimum number of exponentials, 
the fit results are independent of the number of exponentials included in the 
fit. Figure \ref{fig:ds2pt} illustrates the results of this test for the $D_s$ 
meson two-point fits on ensemble set F1. The upper panel presents our 
results for four values of the spatial momentum, plotted as a function of the 
number of exponentials included in the plot. The lower panel shows the 
results obtained from three types of fits: a simultaneous fit to correlator 
data for all four spatial momenta, plotted with blue diamonds; a chained fit 
(discussed in detail in Appendix A of \cite{Bouchard:2014ypa}) to 
correlator data for all four spatial momenta simultaneously, shown with red 
squares; and an ``individual'' fit, plotted with purple circles. These 
individual fits include the correlator data for just a single daughter 
meson momentum in each fit. 

We take the 
result for $N_{\mathrm{exp}} = 5$ from the chained fit as our final 
result for each momentum. These results are tabulated in 
Table \ref{tab:ds2pt} and shown in Figure \ref{fig:ds2pt} as shaded bands in 
each plot. All three fit approaches give consistent results, as seen in the 
lower panel of Figure \ref{fig:ds2pt}, but the simultaneous fits, with or 
without chaining, have the advantage that they capture the correlations between 
momenta, which is then reflected in the uncertainty quoted in the fit results. 
The chained fits give slightly better values of reduced $\chi^2$. For example, 
for the ground state results plotted in the lower panel, the chained fits give 
$\chi^2/\mathrm{dof} = 0.88$ for $N_{\mathrm{nexp}} = 5$, while the 
simultaneous 
fits give $\chi^2/\mathrm{dof} = 1.1$. Both fits include 164 degrees of 
freedom. 
In addition, the chained fits are about ten percent faster than the 
simultaneous 
fits---14.6s to generate all the data in the lower plot for the chained fit 
compared to 16.4s for the simultaneous fit. This is not an important 
consideration for the two-point fits, but becomes relevant for the larger 
three-point fits, which can take many hours. Choosing to use chained fits for 
both two- and three-point fits ensures a consistent approach throughout the 
fitting procedure.
\begin{table}
\caption{\label{tab:ds2pt}
Fit results for the ground state energies of the $D_s$ meson at each spatial 
momentum $\vec{p}$. We take $N_{\mathrm{exp}} = 5$ and fit all 
two-point correlator data simultaneously.}
\begin{ruledtabular}
\begin{tabular}{ccccc}
Set & $a M_{D_s}$   & $a E_{D_s}(1,0,0)$ & $a E_{D_s}(1,1,0)$ & $  
a E_{D_s}(1,1,1)$  \\
\vspace*{-10pt}\\
\hline 
\vspace*{-6pt}\\
C1 & 1.18755(22) & 1.21517(34) & 1.24284(33) & 1.27013(39) \\
C2 & 1.20090(30) & 1.24013(56) & 1.27822(61) & 1.31543(97) \\
C3 & 1.19010(33) & 1.23026(53) & 1.26948(54) & 1.30755(79) \\
F1 & 0.84674(12) & 0.87559(19) & 0.90373(20) & 0.93096(26) \\
F2 & 0.84415(14) & 0.87348(25) & 0.90145(25) & 0.92869(33) \\
\end{tabular}
\end{ruledtabular}
\end{table}

\begin{figure}
\centering
\caption{\label{fig:dsdispersion}Dispersion relation
for each ensemble. The shaded region 
corresponds to $1\pm\alpha_s(ap/\pi)^2$ where we take $\alpha_s=0.25$.}
\includegraphics[width=0.48\textwidth,keepaspectratio]{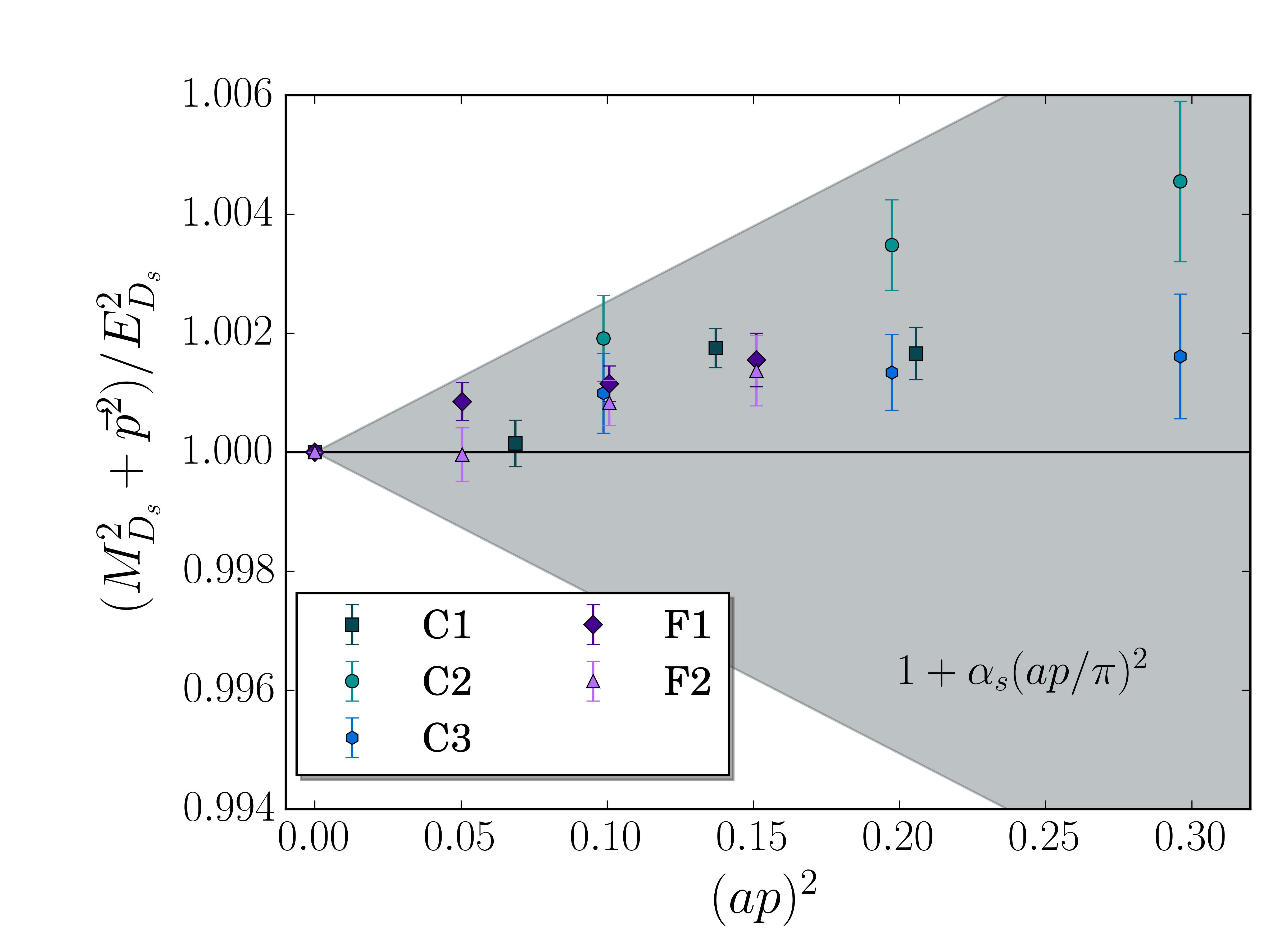}
\end{figure}

As a further test of the two-point fits for the $D_s$ meson we determine the 
ratio $(M_{D_s}^2+\vec{p}^2)/E_{D_s}^2$ on each ensemble. We plot the results 
in Figure \ref{fig:dsdispersion}. The shaded region corresponds to $1\pm 
\alpha_s(ap/\pi)^2$, where we set $\alpha_s = 0.25$. In general, the data
lie systematically above the relativistic value of unity, indicating that the 
statistical uncertainties of the fit results are sufficiently small that we 
can resolve discretization effects at ${\cal O}(\alpha_s(ap/\pi)^2)$. 
These discretization effects are less than $0.5\%$ in the dispersion relation.

\begin{figure}
\centering
\caption{\label{fig:bs2pt}Fit results for the $B_s$ meson two-point correlator 
as a function of the number of exponentials included in the fit on two 
ensemble sets, C2 and F1. We plot our final results, for 
which $N_{\mathrm{exp}}=5$, as a green hexagon for C2 and a purple square for 
F1, with corresponding shaded bands.}
\includegraphics[width=0.48\textwidth,keepaspectratio]{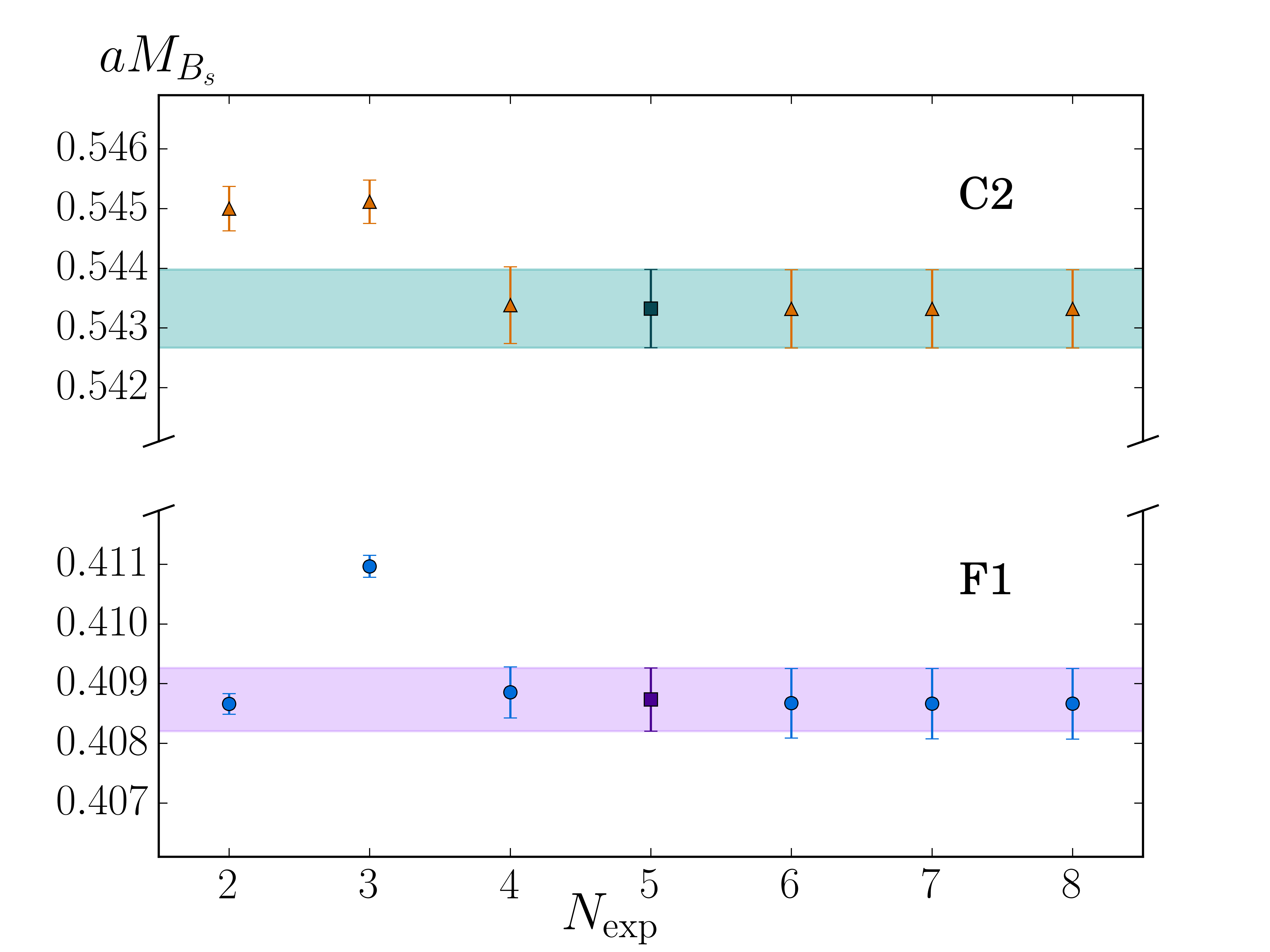}
\end{figure}

Figure \ref{fig:bs2pt} shows the corresponding two-point fit results for the 
ground state of the $B_s$ meson for ensemble sets C2 and F1. These ensemble 
sets have the same sea quark mass ratios, $m_\ell/m_s = 1/5$ (see Table 
\ref{tab:milc}) and the difference between the results stems almost entirely 
from the lattice spacing. We take the values with $N_{\mathrm{exp}} = 
5$ as our final results, highlighted in the figure by the square data points 
and the shaded bands. We tabulate 
our final results in Table \ref{tab:bs2pt}.
\begin{table}
\caption{\label{tab:bs2pt}
Fit results for the ground state $aE_0^{B_s,\mathrm{sim}}$, on each 
ensemble set, with $N_{\mathrm{exp}}=5$.}
\begin{ruledtabular}
\begin{tabular}{ccccc}
C1 & C2 & C3 & F1 & F2 \\
\vspace*{-10pt}\\
\hline 
\vspace*{-7pt}\\
0.53714(60) & 0.54332(65) & 0.53657(86) & 0.40873(53) & 0.40819(44) \\
\end{tabular}
\end{ruledtabular}
\end{table}

For the three-point correlator fits, we use a fitting procedure that diverges 
slightly from the approach taken in 
\cite{Na:2015kha} and do not employ a ``mixed'' fitting strategy. Instead of 
combining ``individual'' and ``master'' fits (see \cite{Na:2015kha} for full 
details), we use chained fits to correlators at all spatial momenta. 
This fitting approach ensures that we keep track of all statistical 
correlations between data at different momenta while maintaining fit stability, 
which was an issue for simultaneous fits attempted in \cite{Na:2015kha}.

\begin{figure}
\centering
\caption{\label{fig:3ptexp}Fit results for the three-point amplitudes
as a function of the number of exponentials on two 
ensemble sets, C2 and F1. We fit to correlator data for all 
values of the spatial momentum simultaneously and thin by keeping every third 
timeslice. We plot our final results, for 
which $N_{\mathrm{exp}}=5$, as a green hexagon for C2 and a purple square for 
F1, with corresponding shaded bands. Note that the amplitudes on set C2 
are approximately three times larger than the amplitudes on set F1, as 
indicated by the left (F1) and right (C2) vertical axes.}
\includegraphics[width=0.48\textwidth,keepaspectratio]{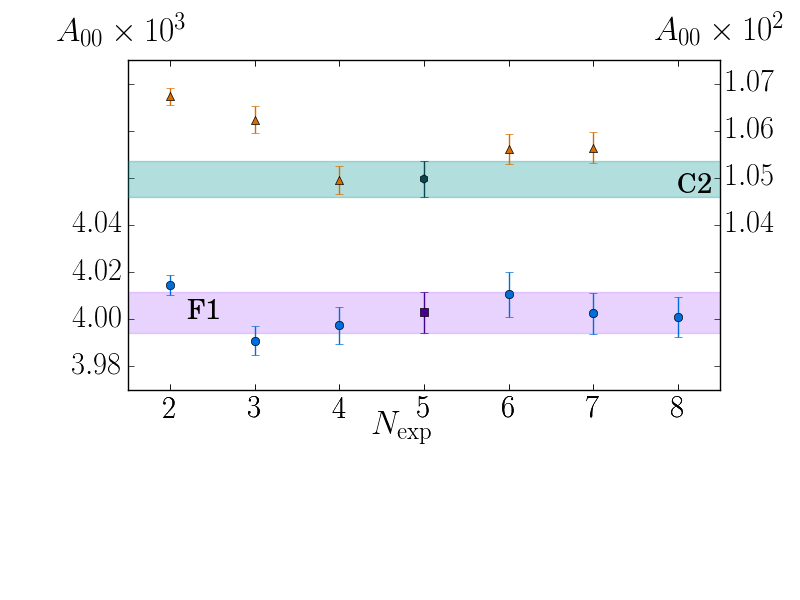}
\end{figure}

To improve stability and goodness-of-fit, we thin the 
three-point correlator data on the fine ensembles by keeping every third 
timeslice. We illustrate the stability of these fits with the number of 
exponentials in the fit in Figure \ref{fig:3ptexp}.

\begin{figure}
\centering
\caption{\label{fig:3ptthin}Fit results for the three-point amplitudes
as a function of the number of exponentials for different choices of data 
thinning: no thinning, represented by turquoise triangles; keeping every third 
timeslice, represented by blue circles and the label ``Thinning = 3''; and 
every fifth timeslice, shown by yellow pentagons and the label ``Thinning = 
5''. Our final result, for which we use thinning by every third timeslice and 
$N_{\mathrm{exp}}=5$, is shown as a purple square and the corresponding purple 
shaded band.}
\includegraphics[width=0.48\textwidth,keepaspectratio]{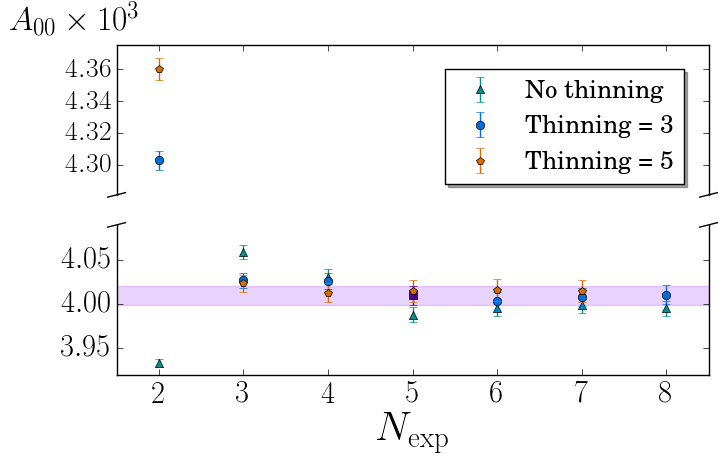}
\end{figure}

We test our 
choice by comparing fit results for the three-point amplitudes with thinning 
(keeping both every third and every fifth timeslice) and without thinning and 
plot the results in Figure \ref{fig:3ptthin}. We do not consider thinning by 
an even integer, which removes information about the oscillatory states 
generated by the staggered 
quark action.

\begin{figure}
\centering
\caption{\label{fig:3ptT}Fit results for the three-point amplitude $A_{00}$
as a function of the number of source-sink separations, $T$, incorporated in 
the fit on ensemble set F1. We fit to correlator data for all 
values of the spatial momentum simultaneously and thin by keeping every third 
timeslice. For our final results we take the full set, $T = (12,13,14,15)$ on 
the coarse ensembles and $T = (21,22,23,24)$ on the fine ensembles, indicated 
by the first point, the purple square, and the purple shaded band. Fit 
results from other combinations of source-sink separations are plotted as 
blue circles.}
\includegraphics[width=0.48\textwidth,keepaspectratio]{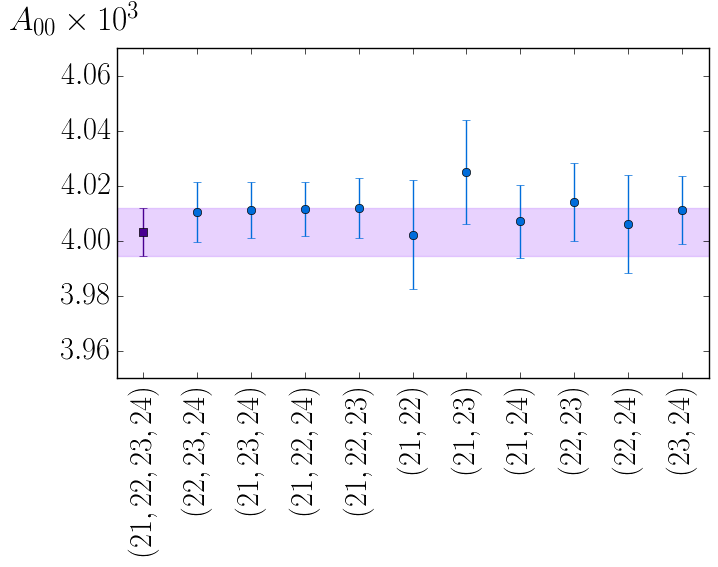}
\end{figure}

In Figure \ref{fig:3ptT} we present results for the three-point fits when 
different combinations of source-sink separations, $T$, are used. For our final 
results we take the full set, $T = (12,13,14,15)$ on the coarse ensembles and 
$T = (21,22,23,24)$ on the fine ensembles.
We fit the three-point correlator data after matching the bottom-charm currents 
to full QCD, as described briefly in Section \ref{sec:lattsetup} and in more 
detail in \cite{Na:2015kha}. In \cite{Na:2015kha} this approach was compared 
with fitting the data first and then matching to full QCD and, as expected, the 
results are in good agreement within errors.

\begin{figure}
\centering
\caption{\label{fig:heatmap}Correlations between form factors at different 
momenta for the ensemble set F2.}
\includegraphics[width=0.48\textwidth,keepaspectratio]{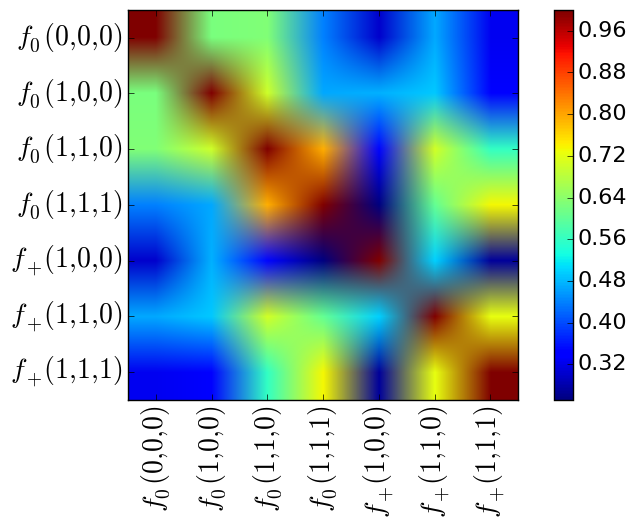}
\end{figure}

We summarize our final results for the form factors, $f_0(\vec{p})$ and 
$f_+(\vec{p})$, for each ensemble and $D_s$ momentum in Tables 
\ref{tab:f0} and 
\ref{tab:fp}. We 
represent the correlations between form factors at different momenta as a heat 
map in Figure \ref{fig:heatmap} for 
ensemble set F2.
\begin{table}
\caption{\label{tab:f0}
Final results for the form factor $f_0(\vec{p})$.}
\begin{ruledtabular}
\begin{tabular}{ccccc}
Set & $f_0(0,0,0)$   & $f_0(1,0,0)$ & $f_0(1,1,0)$ & $  
f_0(1,1,1)$  \\
\vspace*{-10pt}\\
\hline 
\vspace*{-8pt}\\
C1 & 0.8885(11) & 0.8754(14) & 0.8645(13) & 0.8568(13) \\
C2 & 0.8822(13) & 0.8663(15) & 0.8524(16) & 0.8418(18) \\
C3 & 0.8883(13) & 0.8723(16) & 0.8603(16) & 0.8484(21) \\
F1 & 0.90632(98) & 0.8848(13) & 0.8674(13) & 0.8506(17) \\
F2 & 0.9047(12) & 0.8855(16) & 0.8667(15) & 0.8487(19) \\
\end{tabular}
\end{ruledtabular}
\caption{\label{tab:fp}
Final results for the form factor $f_+(\vec{p})$.}
\begin{ruledtabular}
\begin{tabular}{cccc}
Set & $f_+(1,0,0)$   & $f_+(1,1,0)$ & $f_+(1,1,1)$ \\
\vspace*{-10pt}\\
\hline 
\vspace*{-8pt}\\
C1 & 1.1384(35) & 1.1081(20) & 1.0827(21) \\
C2 & 1.1137(29) & 1.0795(22) & 1.0470(21) \\
C3 & 1.1260(34) & 1.0912(24) & 1.0552(28) \\
F1 & 1.1453(29) & 1.0955(24) & 1.0549(24) \\
F2 & 1.1347(42) & 1.0905(26) & 1.0457(33) \\
\end{tabular}
\end{ruledtabular}
\end{table}

\section{\label{sec:zexp}Chiral, continuum and kinematic extrapolations}

\begin{figure}
\centering
\caption{\label{fig:ffbandbs}Form factor results for the $B_s\to D_s\ell\nu$ 
decay, compared to those for the 
$B\to D\ell\nu$ decay from \cite{Na:2015kha}, as function of $z$. We plot four 
sets of results, for $f_0(q^2(z))$ and $f_+(q^2(z))$ for both $B$ and $B_s$ 
meson decays. We distinguish the data in four ways. First, the shape of each 
data marker indicates the corresponding ensemble set, as shown in the legend in 
the upper left corner: squares represent set C1; diamonds set C2; circles C3; 
left-triangles F1; and triangles F2. Second, the upper set of points are those 
for 
$f_+(q^2(z))$ and the lower set of points show the data for $f_0(q^2(z))$, as 
indicated by the annotations. Third, the color of the points distinguishes 
the data as follows: the turquoise-green points represent $f_+^{B_s\to 
D_s}(q^2(z))$; the light purple points are $f_+^{B\to D}(q^2(z))$; the blue 
points are $f_0^{B_s\to D_s}(q^2(z))$; and the orange-yellow
points are $f_0^{B\to D}(q^2(z))$. Finally, we distinguish the data by size:
the larger markers represent the $B\to D\ell\nu$ decay, while the smaller 
points are from those for the $B_s\to D_s\ell\nu$ decay.}
\includegraphics[width=0.48\textwidth,keepaspectratio]{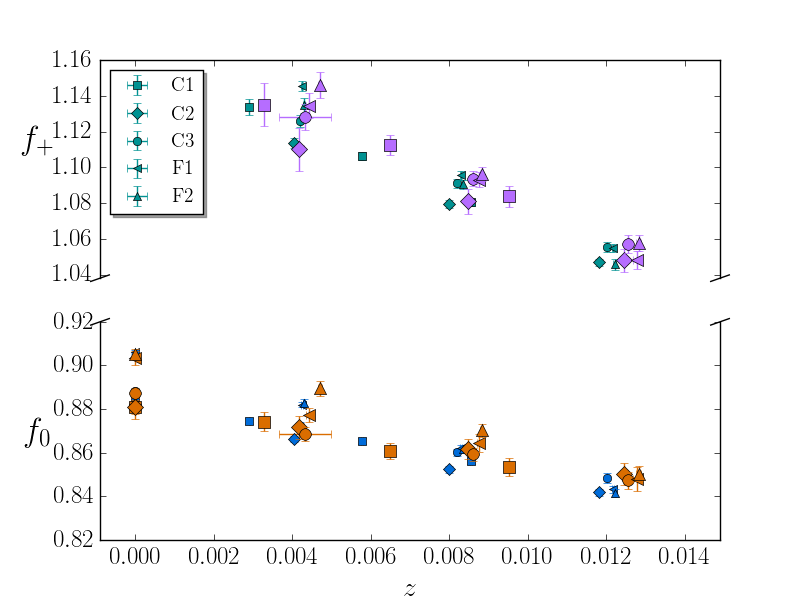}
\end{figure}

The form factor results presented in the previous section are determined at 
finite lattice spacing, with sea quark masses that are heavier than their
physical values. These form factors are therefore functions of the momentum 
transfer, the lattice spacing, and the sea quark masses. The form factors 
determined from experimental data are functions of a single 
kinematic variable only. Typically this variable is the momentum transfer, 
$q^2$, or the daughter meson energy, $E_{D_s}$, but the form factors can also 
be expressed in terms of the $w$-variable, defined by
\begin{equation}\label{eq:wdef}
w(q^2) = 1 + \frac{q_{\mathrm{max}}^2-q^2}{2M_{B_s}M_{D_s}},
\end{equation}
where $q_{\mathrm{max}}^2 = (M_{B_s}-M_{D_s})^2\simeq \SI{11.54}{GeV^2}$ or the 
$z$-variable,
\begin{equation}
z(q^2) = \frac{\sqrt{t_+-q^2} - \sqrt{t_+-t_0}}{\sqrt{t_+-q^2}+\sqrt{t_+-t_0}}.
\end{equation}
Here $t_+ = (M_{B_s}+M_{D_s})^2$ and $t_0$ is a free parameter, which we take 
to be $t_0 = q_{\mathrm{max}}^2$ to ensure consistency with the analysis of 
\cite{Na:2015kha}. In Figure 
\ref{fig:ffbandbs} we compare our 
results for the form factors, 
$f_0(q^2)$ and $f_+(q^2)$, with the corresponding form factors for the $B\to 
D\ell\nu$ decay, taken from \cite{Na:2015kha}, as a function of the 
$z$-variable. From the plot, we see that 
there is little dependence on the light spectator quark species in the form 
factor results.

To relate the form factor results determined at finite lattice spacing and 
unphysical sea quark masses to experimental data, we must therefore perform 
continuum and chiral extrapolations, along with a kinematic extrapolation in 
terms of one of the choices of kinematic variable. We combine these 
extrapolations through the modified $z$-expansion, introduced in 
\cite{Na:2010uf,Na:2011mc}, and applied to $B_{(s)}$ heavy-light decays in 
\cite{Bouchard:2013mia,Bouchard:2013pna,Bouchard:2014ypa}. Our analysis of the 
chiral-continuum-kinematic extrapolation for $B_s \to D_s \ell \nu$ decay 
closely parallels that for the $B \to D \ell \nu$ decay in \cite{Na:2015kha}, 
so we only briefly outline the key components and refer the reader to 
\cite{Na:2015kha} for details.

We express the dependence of the form factors on the $z$-variable through a 
modification of the BCL 
parameterization \cite{Bourrely:2008za}
\begin{align}
f_0(q^2(z)) = {} & \frac{1}{P_0}\sum_{j=0}^{J-1} 
a_j^{(0)}(m_l,m_l^{\mathrm{sea}},a) 
z^j, \\
f_+(q^2(z)) = {} & \frac{1}{P_+}\sum_{j=0}^{J-1} 
a_j^{(+)}(m_l,m_l^{\mathrm{sea}},a)\nonumber\\
{} & \qquad \times
\left[z^j - (-1)^{j-J}\frac{j}{J}z^J\right].
\end{align}
Here the $P_{0,+}$ are Blaschke factors that take into account the effects of 
expected poles above the physical region, 
\begin{equation}
P_{0,+}(q^2) = \left(1-\frac{q^2}{M_{0,+}^2}\right),
\end{equation}
where we take $M_+ = M_{B_c^\ast} = \SI{6.330(9)}{GeV}$ \cite{Gregory:2009hq}, 
and $M_0 = \SI{6.42(10)}{GeV}$. We find little dependence on the value of 
$M_0$, in 
line with the results of \cite{Na:2015kha}. The expansion coefficients 
$a_j^{(0,+)}$ include 
lattice spacing and light quark mass dependence and can be written as
\begin{equation}
a_j^{(0,+)}(m_l,m_l^{\mathrm{sea}},a)  = 
\widetilde{a}_j^{(0,+)}\widetilde{D}_j^{(0,+)}(m_l,m_l^{\mathrm{sea}},a),
\end{equation}
where the $\widetilde{D}_j^{(0,+)}$ include all lattice artifacts and chiral 
logarithms. These coefficients are given by
\begin{align}\label{eq:Dcoeff}
\widetilde{D}_j = {} & 1 + c_j^{(1)} x_\pi+ c_j^{(2)} x_\pi 
\log(x_\pi) \nonumber\\
{} & \qquad + 
d_j^{(1)}\left(\frac{\delta x_\pi}{2} + \delta x_K\right)  + 
d_j^{(2)}\delta x_{\eta_s}
\nonumber\\
{} & \qquad + e_j^{(1)} \left(\frac{aE_{D_s}}{\pi}\right)^2 + e_j^{(2)}
\left(\frac{aE_{D_s}}{\pi}\right)^4 \nonumber\\
{} & \qquad + m_j^{(1)} (am_c)^2 + m_j^{(2)} (am_c)^4,
\end{align}
where
\begin{align}
x_{\pi,K,\eta_s} = {} & \frac{M_{\pi,K,\eta_s}^2}{(4\pi f_\pi)^2}, \\
\delta x_{\pi,K} = {} & \frac{(M_{\pi,K}^{\mathrm{AsqTad}})^2 - 
(M_{\pi,K}^{\mathrm{HISQ}})^2
}{(4\pi f_\pi)^2},  \\
\delta x_{\eta_s} = {} & \frac{(M_{\eta_s}^{\mathrm{HISQ}})^2 - 
(M_{\eta_s}^{\mathrm{phys.}})^2
}{(4\pi f_\pi)^2}, 
\end{align}
and the $c_j^{(i)}$, $d_j^{(i)}$, $e_j^{(i)}$, and $m_j^{(i)}$ are fit 
parameters, 
along with the $\widetilde{a}_j^{(0,+)}$. We use the fit function form of 
\cite{Na:2015kha}, with a new fit parameter, $d_j^{(2)}$, to 
account for the tuning of the valence strange quark mass on each ensemble. We 
tabulate the meson masses required 
to calculate $\delta x_{\pi,K,\eta_s}$ in Table \ref{tab:deltapi}.

We further modify the $z$-expansion parameterization of the form factors to 
accommodate the systematic uncertainty associated with the truncation of the 
matching procedure at ${\cal O}(\alpha_s, \Lambda_{\mathrm{QCD}}/m_b, 
\alpha_s/(am_b))$. We introduce fit parameters $m_\parallel$ and $m_\perp$, 
with central value zero and width $\delta m_{\parallel,\perp}$ and re-scale the 
form factors, $f_\parallel$ and $f_\perp$ according to
\begin{equation}
f_{\parallel,\perp} \rightarrow (1+m_{\parallel,\perp})f_{\parallel,\perp}.
\end{equation}
We take the systematic uncertainties in these fit parameters as 
3\% and refer the reader to the detailed discussion of this approach in 
\cite{Na:2015kha}.

\begin{figure}
\centering
\caption{\label{fig:ffres1}Fit results from the ``standard extrapolation'' fit 
ansatz detailed in the text. The purple data points show the fit results 
at finite lattice spacing and the red and purple shaded bands are the physical 
extrapolations.}
\includegraphics[width=0.48\textwidth,keepaspectratio]{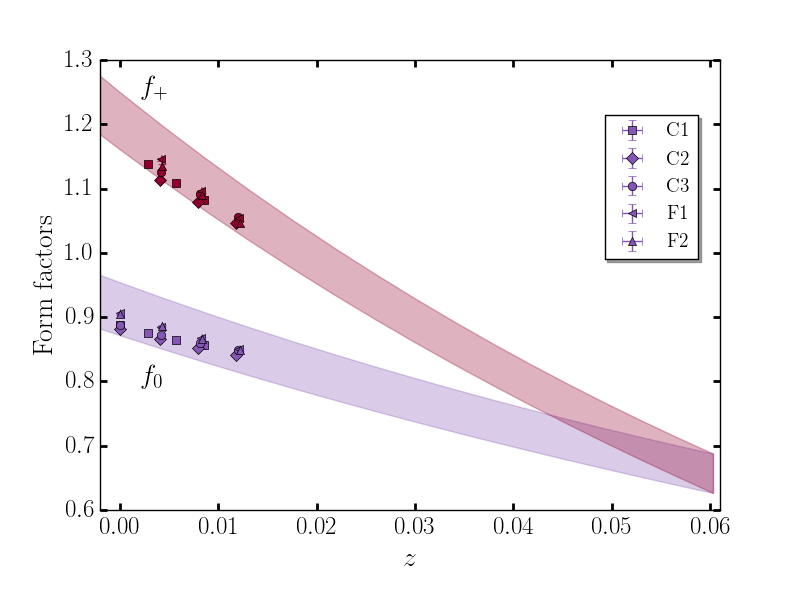}
\end{figure}

\begin{figure}
\centering
\caption{\label{fig:deltaz}Fit results from modifications to the ``standard 
extrapolation'' fit ansatz, plotted as blue circles representing the form 
factor $f_0$ at $q^2=0$ (the lower set of data points) and at $q^2 = 
q^2_{\mathrm{max}}$ (the upper set of points). The test 
numbers labeling the horizontal axis 
correspond to the modifications listed in the text. The first data point, 
the purple square for $f_0(q^2=0)$ and turquoise diamond for $f_0(
q^2_{\mathrm{max}})$, are the 
``standard extrapolation'' 
fit results, which are also represented by the purple and turquoise shaded 
bands, respectively.}
\includegraphics[width=0.48\textwidth,keepaspectratio]{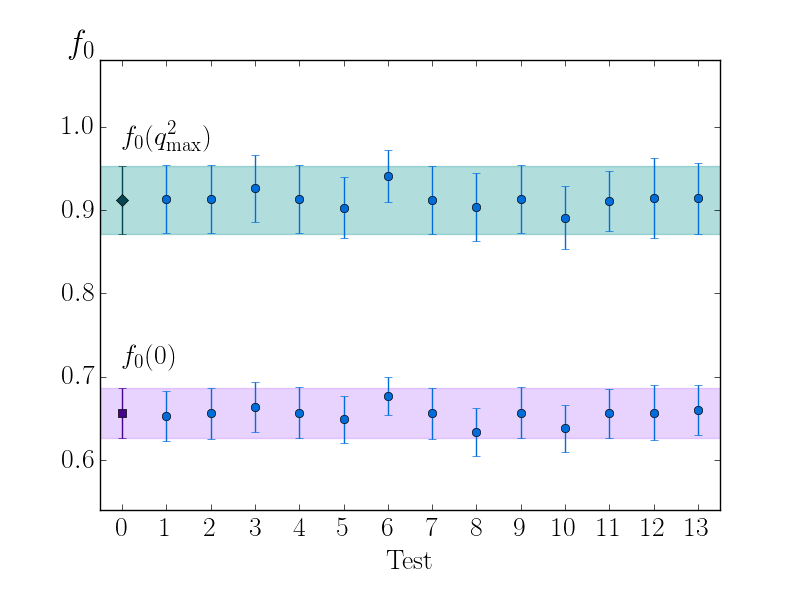}
\end{figure}

In Figure \ref{fig:ffres1}
we plot our fit results for 
$f_0(z)$, $f_+(z)$ as a function of the $z$-variable. We obtain a reduced 
$\chi^2$ of $\chi^2/\mathrm{dof} = 1.2$ with 36 degrees of freedom (dof), with 
a quality factor of $Q=0.24$. The $Q$-value (or $p$-value) corresponds to the 
probability that the $\chi^2/\mathrm{dof}$ from the fit could have been larger, 
by chance, assuming the data are all Gaussian and consistent with each other. 
We plot the lattice data 
and the results of the chiral-continuum-kinematic extrapolation for 
$f_+(z)$ as the upper, red shaded band and for $f_0(z)$ as the lower, purple
shaded band. We use the fit ansatz outlined above, including terms 
up to $z^3$ in the modified $z$-expansion, and refer to these results as the 
``standard extrapolation''. We tabulate our choice 
of priors and the fit results in Appendix \ref{sec:ffdetails}, and provide the 
corresponding $z$-expansion coefficients and their correlations in Table 
\ref{tab:Bszexp}. Following \cite{Na:2015kha} and the 
earlier work of \cite{Na:2010uf,Na:2011mc}, we group the priors into 
Group I and Group II variables, and add a third group. Broadly 
speaking, Group I priors are the 
typical fit parameters, Group II includes the input lattice scales and masses, 
and Group III priors are physical input masses. 
See the appendix of \cite{Na:2015kha} for more details.
To test the convergence of our fit ansatz, we follow a procedure similar to 
that outlined in \cite{Na:2015kha}. This can be summarized 
as modifying the fit ansatz in the following ways:
\begin{enumerate}
\item include terms up to $z^2$ in the $z$-expansion;
\item include terms up to $z^4$ in the $z$-expansion;
\item add light-quark mass dependence to the fit parameters $m_j^{(i)}$;
\item add strange-quark mass dependence to the fit parameters $m_j^{(i)}$;
\item add bottom-quark mass dependence to the fit parameters $m_j^{(i)}$;
\item include discretization terms up to $(am_c)^2$;
\item include discretization terms up to $(am_c)^6$;
\item include discretization terms up to $(aE_{D_s}/\pi)^2$;
\item include discretization terms up to $(aE_{D_s}/\pi)^6$;
\item omit the $x_\pi \log(x_\pi)$ term;
\item incorporate a 2\% uncertainty for higher-order matching contributions;
\item incorporate a 4\% uncertainty for higher-order matching contributions;
\item incorporate 4\% and 2\% uncertainties on coarse and fine ensembles, 
respectively, for higher-order matching contributions.
\end{enumerate}
We show the results of these modifications in Figure \ref{fig:deltaz}. This 
plot demonstrates that the fit has converged with respect to a 
variety of modifications of the chiral-continuum-kinematic extrapolation 
ansatz. As part of this process, we also tested the significance of the 
Blaschke factor in the fit results. In line with the results of 
\cite{Na:2015kha}, we found that, while the results agreed within 
uncertainties, removing the Blaschke lowered the central value and increased 
the uncertainty of the result. This test is not strictly a test of convergence 
and is therefore not included in Figure \ref{fig:deltaz}.

\begin{figure}
\centering
\caption{\label{fig:bbsheatmap}Correlations between $B\to D\ell\nu$ and $B_s\to 
D_s\ell\nu$ ensemble-averaged, three-point correlators for ensemble set C1. The 
data correspond to a single $B_{(s)}$ meson source with Gaussian smearing 
$r_0/a=5$, a source-sink separation of $T = 13$ and with $a\vec{p}_{D_{(s)}} = 
(0,0,0)$.}
\includegraphics[width=0.5\textwidth,keepaspectratio]{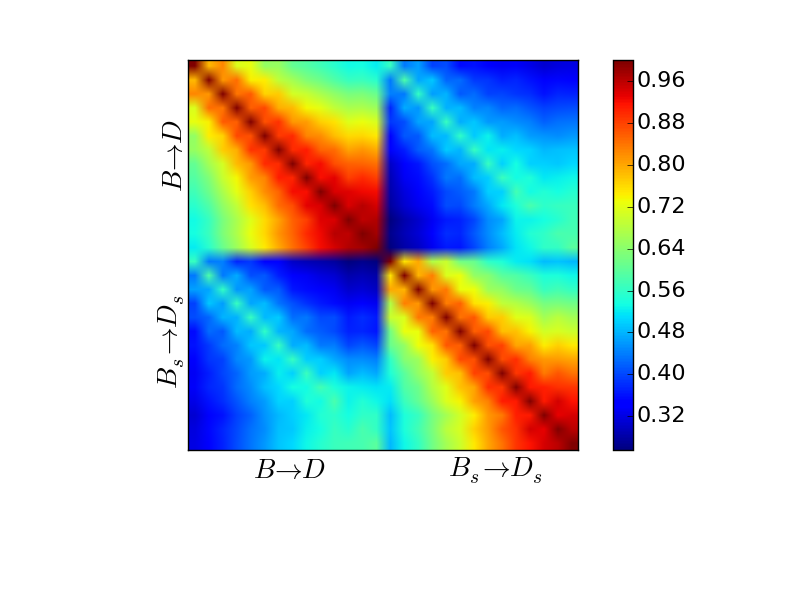}
\end{figure}

To determine the ratio of form factors, we 
simultaneously fit the lattice form factor data for the $B_s\to D_s\ell\nu$ and 
$B\to D\ell\nu$ decays in a single script. We take the form factor results from 
Table III of \cite{Na:2015kha} for the $B\to D\ell\nu$ decay. Fitting the 
results simultaneously ensures that statistical correlations between the two 
data sets, such as those stemming from the lattice spacing determination on 
each ensemble set, are included in the final result for the ratio at zero 
momentum transfer. We do not re-analyze the $B\to D\ell\nu$ to account for 
statistical correlations between the correlators themselves, which have 
negligible effect on the final result, given the current precision. This 
analysis would require fitting both $B\to D\ell\nu$ and $B_s\to D_s\ell\nu$ 
two- and three-point correlators simultaneously. To ensure that these 
statistical correlations are not important, we tested the correlations 
between the three-point correlators on different ensemble sets. We show an 
example of the corresponding correlations as a heat map in Figure 
\ref{fig:bbsheatmap}, from which one can see that statistical correlations are 
less than $\sim 0.6$. We have found that correlations of this size have 
negligible impact at our current level of precision.

We fit the form factor data using the standard extrapolation ans\"atze for both 
the $B\to D\ell\nu$ and $B_s\to D_s\ell\nu$ data. For the $B_s\to D_s\ell\nu$ 
decay, we choose the priors for the coefficients in the modified $z$-expansion 
to be equal to those for the corresponding expression for the $B\to D\ell\nu$ 
$z$-expansion. These priors reflect the close agreement 
between the values for the $B\to D\ell\nu$ and $B_s\to D_s\ell\nu$ decays, 
illustrated in Figure \ref{fig:ffbandbs}. We list our choice 
of priors and the fit results for the ratio of form factors in Appendix 
\ref{sec:ffdetails}, and provide the 
corresponding $z$-expansion coefficients and their correlations in Table 
\ref{tab:BBszexp}.

\section{\label{sec:results}Results}

\subsection{Form factors}

\begin{figure}
\centering
\caption{\label{fig:ffq2}Chiral and continuum extrapolated form factors, 
$f_0(q^2)$ (lower band) and $f_+(q^2)$ (upper band), as a function of the 
momentum transfer.}
\includegraphics[width=0.48\textwidth,keepaspectratio]{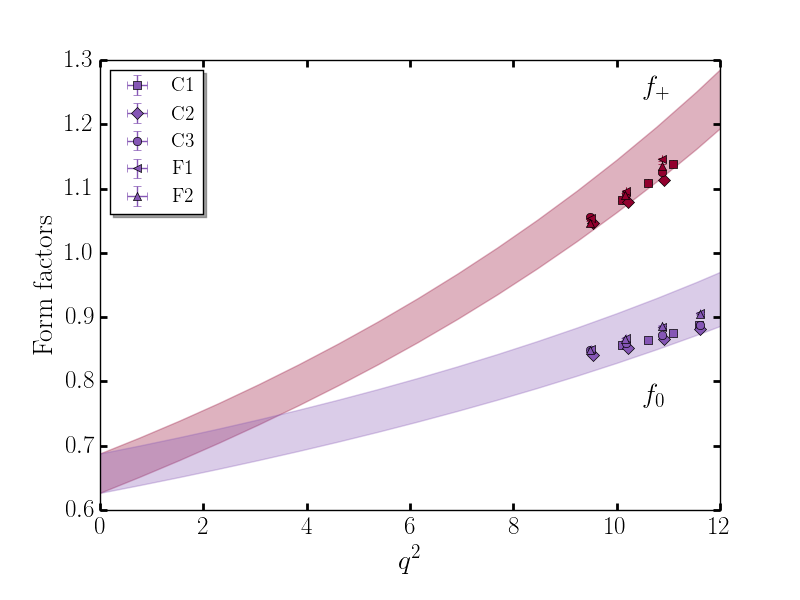}
\end{figure}

We plot our final results for the form factors, $f_0(q^2)$ and $f_+(q^2)$, as a 
function of the momentum transfer, $q^2$, in Figure \ref{fig:ffq2}.

\begin{figure}
\centering
\caption{\label{fig:ffratioq2}Chiral and continuum extrapolated form factors, 
$f_0(q^2)$ (lower band) and $f_+(q^2)$ (upper band), as a function of the 
momentum transfer, for both $B_s\to D_s$ (purple hatched band) and $B \to D$ 
(plain turquoise band) semileptonic decays. The lattice data for each 
decay cannot be distinguished on this plot and are therefore not included. See 
Figure \ref{fig:ffres1} for a detailed plot of the results for the form factors 
at finite lattice spacing for both decays.}
\includegraphics[width=0.48\textwidth,keepaspectratio]{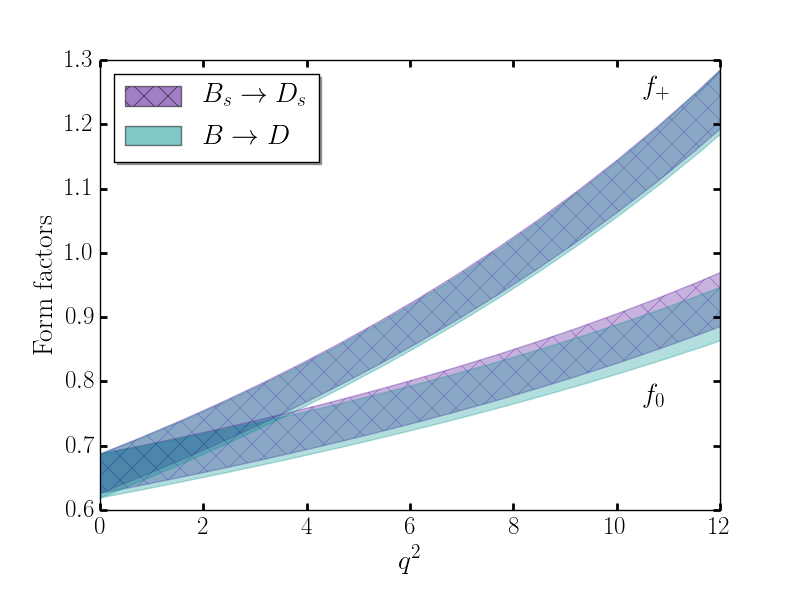}
\end{figure}

Our final result for the form factor at zero momentum transfer is
\begin{equation}\label{eq:ff0result}
f_0^{B_s\to D_s}(0) = f_+^{B_s\to D_s}(0) = 0.656(31).
\end{equation}
We provide an estimate of the error budget for this result in Table 
\ref{tab:f0errors}.
For the ratio of form factors, we find
\begin{equation}\label{eq:ffKresult}
\frac{f_0^{B_s\to D_s}(M_\pi^2)}{f_0^{B\to D}(M_K^2)}  = 
1.000(62),
\end{equation}
and
\begin{equation}\label{eq:ffmpiresult}
\frac{f_0^{B_s\to D_s}(M_\pi^2)}{f_0^{B\to D}(M_\pi^2)}  = 
1.006(62),
\end{equation}
with corresponding error budgets in Table \ref{tab:fratioerrors}.
We show the extrapolation bands as a 
function of momentum transfer for both $B_s\to D_s$ (purple hatched band) and $B 
\to D$ 
(plain turquoise band) semileptonic decays in Figure \ref{fig:ffratioq2}.

We find agreement, within errors, with the results of 
\cite{Bailey:2012rr}, which are
\begin{align}
\frac{f_0^{B_s\to D_s}(M_\pi^2)}{f_0^{B\to 
D}(M_K^2)}[\mathrm{FNAL/MILC}] = 
{} & 1.046(46)\\ 
\frac{f_0^{B_s\to D_s}(M_\pi^2)}{f_0^{B\to D}(M_\pi^2)}[\mathrm{FNAL/MILC}] = 
{} & 1.054(50).
\end{align}
Here we have combined the uncertainties quoted in \cite{Bailey:2012rr}, which 
are statistical and systematic, in quadrature.

For the form factor at zero recoil, $f_+(q^2_{\mathrm{max}})$, which is often 
quoted as
\begin{equation}
{\cal G}(1) = \frac{2\sqrt{\kappa}}{1+\kappa}f_+(q_{\mathrm{max}}^2),
\end{equation}
where $\kappa = M_{D_s}/M_{B_s}$, we find
\begin{equation}\label{eq:G1result}
{\cal G}(1)  = 1.068(40).
\end{equation}
This result is in good agreement with the value of ${\cal G}(1) = 
1.052(46)$ determined in \cite{Atoui:2013zza}, with a slightly smaller 
uncertainty. The corresponding 
values for the $B\to D\ell\nu$ form factors are ${\cal 
G}^{B\to D}(1) = 1.035(40)$ \cite{Na:2015kha} and ${\cal 
G}^{B\to D}(1) = 1.058(9)$ \cite{Bailey:2012rr} (where the quoted uncertainty 
includes only statistical uncertainties).

The slope of the form factor, $f_+(q^2)$, is 
given by
\begin{equation}
\rho^2(w) = -\frac{{\cal G}'(w)}{{\cal G}(w)},
\end{equation}
where the derivative is with respect to the $w$-variable of Equation 
\eqref{eq:wdef}.
In the CLN parameterization, \cite{Caprini:1997mu}, the form factor is then
parameterized by
\begin{equation}
{\cal G}(w) = {\cal G}(1)\Big[1-8\rho^2z + (51 \rho^2-10)z^2 - (252 
\rho^2-84)z^3\Big],
\end{equation}
with $z = z(w)$ the $z$-variable of the previous section:
\begin{equation}
z(w) = \frac{\sqrt{w+1}-\sqrt{2}}{\sqrt{w+1}+\sqrt{2}}.
\end{equation}
We obtain 
\begin{equation}
\rho^2(1) = 1.244(76)
\end{equation}
for the slope of the form factor.

Experimental data for the $B\to D\ell\nu$ decay is typically 
presented in the form $|V_{cb}|{\cal G}(1)$, since the differential decay 
rate for the $B_{(s)}\to D_{(s)}\ell\nu$ decay can be written as
\begin{align}
\frac{d\Gamma(B_{(s)}\to D_{(s)}\ell\nu)}{dw} = {} & \frac{G_F^2}{48 \pi^3} 
M_{D_{(s)}}^3(M_{B_{(s)}}+M_{D_{(s)}})^2 \nonumber\\
{} & \times (w^2-1)^{3/2}|V_{cb}|^2|{\cal G}(w)|^2,
\end{align}
where $G_F$ is the Fermi constant. In this form, lattice results for the form 
factor ${\cal G}(1)$ provide the normalization required to extract $|V_{cb}|$ 
from 
experimental data. Incorporating the slope of the form factor, $\rho^2(w)$, 
helps further tighten experimental determinations of $|V_{cb}|$. An even more 
powerful approach incorporates the full kinematic dependence on the scalar and 
vector form factors, in combination with experimental data over a range of 
momentum transfer \cite{Na:2015kha,Huschle:2015rga}. When combined with our form 
factor results, future experimental data for the $B_s\to D_s\ell\nu$ decay will 
provide a new 
method to extract $|V_{cb}|$ and may shed light on the long-standing tension 
between exclusive and inclusive determinations of $|V_{cb}|$.

\subsection{Form factor error budget}

We tabulate 
the errors in the form factors at zero momentum transfer, Equation 
\eqref{eq:ff0result}, in Table \ref{tab:f0errors}.
\begin{table}
\caption{\label{tab:f0errors}
Error budget for the form factors at zero momentum transfer, $f_0(0) = 
f_+(0)$, for the $B_s\to D_s \ell\nu$ semileptonic decay. 
We describe each source of uncertainty in more detail in the accompanying text.
}
\begin{ruledtabular}
\begin{tabular}{cc}
Type & Partial uncertainty (\%) \\
\vspace*{-10pt}\\
\hline 
\vspace*{-8pt}\\
Statistical & 1.22 \\
Chiral extrapolation & 0.80 \\
Quark mass tuning & 0.66 \\
Discretization & 2.47 \\
Kinematic & 0.71 \\
Matching & 2.21 \\
\vspace*{-10pt}\\
\hline 
\vspace*{-8pt}\\
total & 3.70 
\end{tabular}
\end{ruledtabular}
%\end{center}
\end{table}
The sources of uncertainty listed in Table \ref{tab:f0errors} are:
\paragraph{Statistical} The statistical uncertainties include the two- and 
three-point correlator fit errors and those associated with the lattice spacing 
determination, $r_1$ and $r_1/a$.
\paragraph{Chiral extrapolation} This uncertainty includes the valence and sea 
quark mass extrapolation errors and chiral logarithms in the chiral-continuum 
extrapolation. These effects correspond to the fit parameters $c_j^{i}$ in 
Equation \eqref{eq:Dcoeff}.
\paragraph{Quark mass tuning} Uncertainties arising from tuning errors in 
the light and strange quark masses at finite lattice spacing, including 
partial quenching effects between the HISQ valence and AsqTad sea quarks. These 
uncertainties are generally very small.
\paragraph{Discretization} Discretization effects incorporate the $(am_c)^n$ 
and $(aE_{D_s}/\pi)^n$ terms in the modified $z$-expansion. These effects are 
the dominant source of uncertainty in our results.
\paragraph{Kinematic} These uncertainties stem from the $z$-expansion 
coefficients and the locations of the poles in the Blaschke factors.
\paragraph{Matching} Matching errors arise from the $m_{\perp,\parallel}$ fit 
parameters discussed in the previous section. Perturbative matching 
uncertainties are the second-largest source of uncertainty in our final results.
We propagate these uncertainties from the large momentum-transfer region, for 
which we have lattice results, to zero momentum-transfer.

The uncertainties associated with physical meson mass input errors and finite 
volume effects, which are both less than $0.01\%$, are not included in these 
estimates, because they are negligible contributions to the final error budget. 
In our error budget, we also neglect uncertainties from electromagnetic 
effects, isospin breaking, and the effects of quenching in the charm quark in 
the gauge ensembles.

In Table \ref{tab:fratioerrors} we list the uncertainties in the form factor 
ratios, Equations \eqref{eq:ffKresult} and \eqref{eq:ffmpiresult}. These 
uncertainties are dominated by those coming from the $B\to D\ell\nu$ decay 
\cite{Na:2015kha}.
\begin{table}
\caption{\label{tab:fratioerrors}
Error budget for the ratio of the form factors, $f_0^{B_s\to 
D_s}(M_\pi^2)/f_0^{B\to D}(M_K^2)$ (second column) and $f_0^{B_s\to 
D_s}(M_\pi^2)/f_0^{B\to D}(M_\pi^2)$ (third column).
We describe each source of uncertainty in more detail in the accompanying text.
}
\begin{ruledtabular}
\begin{tabular}{ccc}
Type & \multicolumn{2}{c}{Partial uncertainty (\%)}{} \\
& $\frac{\vphantom{\big[}f_0^{B_s\to 
D_s}(M_\pi^2)}{\vphantom{\big[}f_0^{B\to D}(M_K^2)}$ & 
$\frac{\vphantom{\big[}f_0^{B_s\to 
D_s}(M_\pi^2)}{\vphantom{\big[}f_0^{B\to D}(M_\pi^2)}$ \\
\vspace*{-8pt}\\
\hline 
\vspace*{-8pt}\\
Statistical & 2.28 & 2.32 \\
Chiral extrapolation & 1.22 & 1.22 \\
Quark mass tuning & 0.81 & 0.81 \\
Discretization & 3.48 & 3.49 \\
Kinematic & 1.38 & 1.43 \\
Matching & 0.07 & 0.05 \\
\vspace*{-10pt}\\
\hline 
\vspace*{-8pt}\\
total & 6.15 & 6.18
\end{tabular}
\end{ruledtabular}
\end{table}

\subsection{Semileptonic decay phenomenology}

With our results for the ratio of the form factors, $f_0^{B_s\to 
D_s}/f_0^{B\to D}$, in Equations \eqref{eq:ffKresult}
and \eqref{eq:ffmpiresult}, we can now determine the ratio of fragmentation 
fractions. LHCb presents their measurement of the these ratios in the form 
\cite{Aaij:2011hi}
\begin{align}
\frac{f_s}{f_d} ={} & 0.310(30)_{\mathrm{stat.}}(21)_{\mathrm{syst.}}
\frac{1}{{\cal N}_a{\cal N}_F},\label{eq:fsfd1}\\
\frac{f_s}{f_d} ={} & 0.307(17)_{\mathrm{stat.}}(23)_{\mathrm{syst.}}
\frac{1}{{\cal N}_a{\cal N}_e{\cal N}'_F},\label{eq:fsfd2}
\end{align}
where the ${\cal N}_a$ parameterize deviations from 
naive factorization and ${\cal N}_e$ is an electroweak correction factor to 
account for $W$-exchange.
The dependence on the form factors is expressed in 
${\cal N}_F$ and ${\cal N}'_F$, which are given in Equation 
\eqref{eq:ffrat}. For convenience, we repeat those expressions here:
\begin{equation}\label{eq:ffrat2}
{\cal N}_F= \left [ \frac{f_0^{(s)}(M_\pi^2)}{f_0^{(d)}(M_K^2)} \right ]^2 
\quad \mathrm{and} \quad 
{\cal N}'_F = \left [ \frac{f_0^{(s)}(M_\pi^2)}{f_0^{(d)}(M_\pi^2)}
 \right ]^2.
\end{equation}
These ratios are relevant to the extraction of the fragmentation fraction 
ratios from the branching fraction ratios
\begin{equation}
\frac{{\cal B}(\overline{B}_s^0\to 
D_s^+\pi^-)}{{\cal B}(\overline{B}^0\to 
D^+K^-)}\quad \mathrm{and} \quad \frac{{\cal B}(\overline{B}_s^0\to 
D_s^+\pi^-)}{{\cal B}(\overline{B}^0\to 
D^+\pi^-)},
\end{equation}
respectively.

Using our results in Equations \eqref{eq:ffKresult}
and \eqref{eq:ffmpiresult}, we obtain
\begin{align}
{\cal N}_F= {} & 1.00(12),\\
{\cal N}'_F ={} & 1.01(12).
\end{align}
These results are uncorrelated with the other factors in Equations 
\eqref{eq:fsfd1} and \eqref{eq:fsfd2}, so that we can update the LHCb 
result for the fragmentation ratio directly. Using the values of ${\cal N}_a = 
1.00(2)$ and ${\cal N}_e = 
0.966(75)$ \cite{Fleischer:2010ay,Fleischer:2010ca}, we find
\begin{equation}\label{eq:fsfdK}
\frac{f_s}{f_d} = 
0.310(30)_{\mathrm{stat.}}(21)_{\mathrm{syst.}}(6)_{theor.}(38)_{\mathrm{latt.}}
\end{equation}
by using ${\cal N}_F$ for the ${\cal B}(\overline{B}_s^0\to 
D_s^+\pi^-)/{\cal B}(\overline{B}^0\to 
D^+K^-)$ channel. The uncertainties in this result are: the experimental 
statistical and systematic uncertainties; the uncertainty associated with
${\cal N}_a$; and the uncertainties in our lattice input, ${\cal N}_F$. We 
assume no correlations in these uncertainties. For the ${\cal 
B}(\overline{B}_s^0\to 
D_s^+\pi^-)/{\cal B}(\overline{B}^0\to 
D^+\pi^-)$ channel, we obtain
\begin{equation}\label{eq:fsfdpi}
\frac{f_s}{f_d} = 
0.307(16)_{\mathrm{stat.}}(21)_{\mathrm{syst.}}(23)_{\mathrm{theor.}}(44)_{
\mathrm{latt.}}
\end{equation}
from ${\cal N}'_F$.

These results are in agreement with the result determined in 
\cite{Bailey:2012rr},
\begin{equation}
\frac{f_s}{f_d} =  
0.286(16)_{\mathrm{stat.}}(21)_{\mathrm{syst.}}(26)_{\mathrm{latt.}}(22)_{
\mathrm {Ne} }.
\end{equation}
Both of these lattice results are a little higher than that quoted in 
\cite{CMS:2014xfa} of $f_s/f_d = 0.259(15)$ or the average value of $f_s/f_d = 
0.267^{+22}_{-20}$ determined in \cite{Aaij:2011jp}, but all results agree 
within the quoted uncertainties.

The ratio
\begin{equation}
R(D) = \frac{{\cal B}(B\to D\tau \nu)}{{\cal B}(B\to D\ell \nu)}
\end{equation}
measures the ratio of branching fraction of the semileptonic decay to the 
$\tau$ lepton to the branching fraction to an electron or muon 
(represented by $\ell$). The experimental measurements of this branching 
fraction ratio are currently in tension with the Standard Model result. 
The global experimental average is 
\cite{Lees:2012xj,Lees:2013uzd,Huschle:2015rga,hfag:2016rds}
\begin{equation}
R(D)_{\mathrm{exp.}} = 0.391(41)_{\mathrm{stat.}}(28)_{\mathrm{sys.}},
\end{equation}
a value that is approximately 4$\sigma$ from the theoretical expectation 
\begin{equation}
R(D)_{\mathrm{theor.}} = 0.299(7),
\end{equation}
where we have taken the mean of the results in
\cite{Kamenik:2008tj,Bailey:2012rr,Na:2015kha}, and combined uncertainties in 
quadrature, neglecting any correlations for simplicity, because a full
analysis of this result is beyond the scope of this work.

\begin{figure}
\centering
\caption{\label{fig:dgdq}Differential branching fractions for the $B_s\to 
D_s\mu \nu$ (hatched magenta band) and $B_s\to D_s \tau\nu$ (purple band) 
decays.}
\includegraphics[width=0.42\textwidth,keepaspectratio]{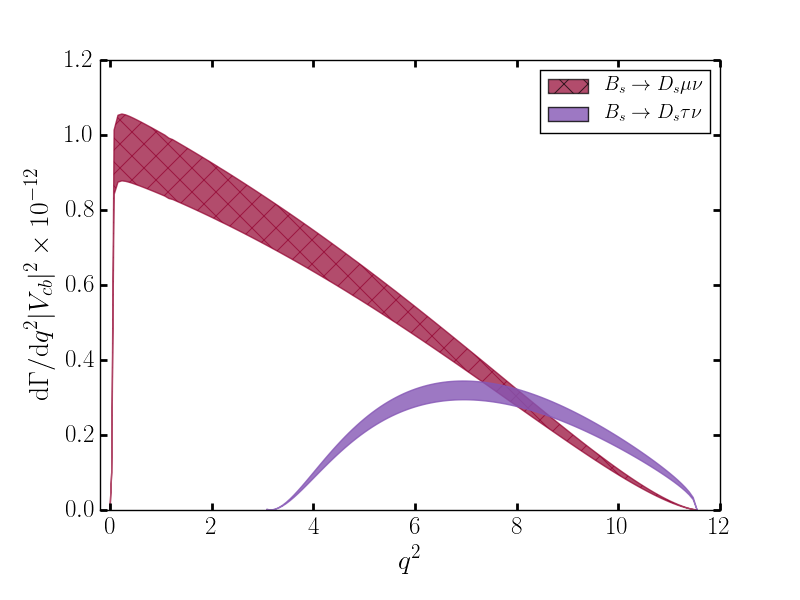}
\end{figure}

We present the first calculation from lattice QCD of the corresponding ratio 
for the semileptonic $B_s\to D_s\ell \nu$ decay, 
\begin{equation}
R(D_s) = \frac{{\cal B}(B_s\to D_s\tau \nu)}{{\cal B}(B_s\to D_s\ell \nu)}.
\end{equation}
This ratio has not been experimentally measured and this provides an 
opportunity for lattice QCD to make a clear prediction of the value expected 
from the Standard Model. Using the form factor results of the previous section, 
we find
\begin{equation}\label{eq:rdsresult}
R(D_s) = 0.314(6).
\end{equation}
We 
provide a complete error budget for this ratio in Table \ref{tab:rdserrors}
and plot the differential branching fractions for $B_s\to D_s\mu \nu$ and 
$B_s\to D_s \tau\nu$ as functions of the momentum transfer in Figure 
\ref{fig:dgdq}.
This result is larger, and about three time more precise, than the prediction 
of $R(D_s) = 0.274^{+20}_{-19}$  
\cite{Bhol:2014jta}, where the form factors were determined from a relativistic 
quark model.
\begin{table}
\caption{\label{tab:rdserrors}
Error budget for the branching fraction ratio $R(D_s)$. 
We describe each source of uncertainty in more detail in the accompanying text. 
The uncertainties associated with discretization effects is no longer the 
dominant source of uncertainty, because the discretization effects largely 
cancel in the ratio.
}
\begin{ruledtabular}
\begin{tabular}{cc}
Type & Partial uncertainty (\%) \\
\vspace*{-10pt}\\
\hline 
\vspace*{-8pt}\\
Statistical & 0.90 \\
Chiral extrapolation & 0.16 \\
Quark mass tuning & 0.19 \\
Discretization & 0.84 \\
Kinematic & 1.13 \\
Matching & 1.05 \\
\vspace*{-10pt}\\
\hline 
\vspace*{-8pt}\\
total &  1.94
\end{tabular}
\end{ruledtabular}
\end{table}

\section{\label{sec:summary}Summary}

We have presented a lattice study of the $B_s \to 
D_s\ell\nu$ semileptonic decay over the full kinematic range of 
momentum transfer and determined
the form factors, $f_0^{B_s \to 
D_s}(q^2)$ and $f_+^{B_s \to D_s}(q^2)$. Combining these results with a 
previous 
determination of the corresponding form factors for the $B\to D\ell\nu$ decay 
\cite{Na:2015kha}, we extracted the ratios $f_0^{B_s \to 
D_s}(M_\pi^2)/f_0^{B \to D}(M_K^2)$ and $f_0^{B_s \to 
D_s}(M_\pi^2)/f_0^{B \to D}(M_\pi^2)$. From these ratios 
we computed the 
fragmentation fraction ratio $f_s/f_d$, an important ingredient in 
experimental determinations of $B_s$ meson branching fractions at hadron 
colliders, particularly for the rare decay ${\cal B}(B_s \rightarrow \mu^+ 
\mu^-)$. In addition, we predict $R(D_s)$, the ratio of the branching fractions 
of the semileptonic $B_s$ decay
to tau and to electrons and muons.

There are a number of tensions between experimental measurements and 
theoretical expectations for semileptonic decays of the $B$ 
meson. These tensions include the branching fraction ratios, $R(D^{(\ast)})$, 
and determinations of $|V_{cb}|$ from exclusive and inclusive decays. Future 
experimental measurements of semileptonic decays of $B_s$ mesons, in 
conjunction with our results for the form factors and for $R(D_s)$, may 
provide some insight into these tensions.

Our result for the form factor at zero recoil, ${\cal G}(1)$, presented in 
Equation \eqref{eq:G1result}, is consistent with an earlier determination by 
the ETM collaboration \cite{Atoui:2013zza}. Moreover, our results for the form 
factor ratios $f_0^{B_s \to D_s}(M_\pi^2)/f_0^{B \to D}(M_K^2)$ and $f_0^{B_s 
\to D_s}(M_\pi^2)/f_0^{B \to D}(M_\pi^2)$, given in Equations 
\eqref{eq:ffKresult} and \eqref{eq:ffmpiresult}, are in 
agreement with the values obtained by the FNAL/MILC 
collaborations. Our determination of this ratio incorporates correlations 
between the form factors for both decay channels, but the 
quoted uncertainty does not include the statistical correlations between 
the raw correlator data, which are negligible at the current level of 
precision. We determine values for the fragmentation fraction ratio, 
$f_s/f_d$, Equations \eqref{eq:fsfdK} and \eqref{eq:fsfdpi}. These results 
have larger uncertainties associated with the form factor inputs than those 
determined in \cite{Bailey:2012rr}.
Finally, we give the branching fraction ratio, 
$R(D_s)$, in Equation \eqref{eq:rdsresult}.

The dominant uncertainty in the form factors for the $B_s \to 
D_s\ell\nu$ decay arises from the discretization effects, with a significant 
contribution from the matching to full QCD. Higher order calculations in 
lattice perturbation theory with the highly 
improved actions employed in this calculation are currently unfeasible, so we 
are exploring ways to reduce matching errors by combining results 
calculated using NRQCD with those determined with an entirely relativistic 
formulation for the $b$-quark. This approach is outlined in 
\cite{Bouchard:2014ypa,Na:2015kha}.

The LHC is scheduled to significantly improve the statistical 
uncertainties in experimental measurements of $B_s$ decays with more data over 
the next decade. Currently, the most precise determinations of the 
fragmentation fraction ratio, $f_s/f_d$, are those measured in situ at the LHC. 
To improve the theoretical calculations of this ratio requires 
several advances. At present the lattice form factor results are the 
largest source of uncertainty in the theoretical result for the ratio, but this 
could be improved with a suitable global averaging procedure, such 
as that undertaken in \cite{Aoki:2016frl}.

Further improvements in the 
uncertainty in the Standard Model expectation of the ratio of the 
fragmentation fractions will ultimately require 
concerted effort to reduce all sources of uncertainty, not just those from 
lattice QCD. Improved theoretical determinations of the fragmentation 
fraction 
ratio will be necessary to take full advantage of the better statistical 
precision of future experimental results and shed light on current
tensions in the heavy quark flavor sector.

\begin{acknowledgments}
Numerical simulations were carried out on facilities
of the USQCD collaboration funded by the Office of Science of the DOE and 
at the Ohio Supercomputer Center. Parts of this work were supported by the 
National Science Foundation. J.S. was supported in part by DOE grant 
DE-SC0011726. C.J.M.~and H.N.~were supported in part by NSF grant 
PHY1414614. We thank the MILC collaboration for use of their gauge 
configurations.
\end{acknowledgments}

\appendix

\section{\label{sec:ffdetails}Reconstructing form factors}

In this appendix we provide our fit results for the coefficients of the 
$z$-expansion, for both the $B_s\to D_s\ell\nu$ decay and the 
ratio of the $B\to D\ell \nu$ and
$B_s\to D_s\ell \nu$ decays. We also 
tabulate our choice of priors for the chiral-continuum extrapolation for the 
$B_s\to D_s\ell\nu$ decay.
\begin{table*}
\caption{\label{tab:Bszexp}
Coefficients of $z$-expansion and the corresponding Blaschke factors (first 
row), and their covariances, for the $B_s\to D_s\ell \nu$ decay. The rows 
correspond to the columns, moving from to bottom and left to right, 
respectively.
}
\begin{ruledtabular}
\begin{tabular}{cccccccc}
$a_0^{(0)}$ & $a_1^{(0)}$ & $a_2^{(0)}$ & $P_0$ & $a_0^{(+)}$ & $a_1^{(+)}$ 
& 
$a_2^{(+)}$ & $P_+$ \\
\vspace*{-8pt}\\
\hline
\vspace*{-8pt}\\
0.658(31) & -0.10(30) & 1.3(2.8) & 6.330(9) & 0.858(32) & -3.38(41) & 
0.6(4.7) & 6.43(10)\\
\vspace*{-10pt}\\
\hline 
\vspace*{-8pt}\\
9.53401$\times 10^{-4}$ & -3.03547$\times 10^{-3}$ &
-5.42391$\times 10^{-3}$ & 8.76501$\times 10^{-4}$  & 5.94503$\times 10^{-4}$ & 
1.58251$\times 10^{-3}$ & 1.60091$\times 10^{-2}$  & 
6.15598$\times 10^{-6}$ \\
  & 9.03097$\times 10^{-2}$  &  -0.101760 & -1.69040$\times 
10^{-2}$  &  4.46248$\times 10^{-4}$
 &   2.36283$\times 10^{-2}$  &  4.56659$\times 10^{-2}$ & -1.29286$\times 
10^{-4}$ \\
&  & 8.02283 & 3.96101$\times 10^{-3}$   &  8.48079$\times 
10^{-3}$
 &   0.104246  &   0.760797 & -8.23960$\times 10^{-7}$ \\
 & & & 1.06275$\times 10^{-2}$ & -3.65165$\times 10^{-5}$ 
&-1.30241$\times 10^{-3}$ & -3.70251$\times 10^{-3}$ &  8.06159$\times 10^{-5}$ 
\\
  &   & &  &   1.00761$\times 10^{-3}$  & -4.23358$\times 
10^{-3}$ 
 &  -2.64511$\times 10^{-2}$ & 9.42502$\times 10^{-6}$ \\
  &   &   &   & & 0.165251  &  -0.617234 & -1.88031$\times 
10^{-4}$\\
  &    &    & & &  & 22.49292 & 6.83236$\times 10^{-5}$\\
 & & & & & & & 8.09911$\times 10^{-5}$ \\
\end{tabular}
\end{ruledtabular}
\end{table*}

\begin{table}
\caption{\label{tab:BBszexp}
Coefficients and Blaschke factors for the $z$-expansions for the 
ratio of the $B_s\to D_s\ell \nu$ and $B\to 
D\ell \nu$, decays. Note that the Blaschke 
factors are common to both expansions.
}
\begin{ruledtabular}
\begin{tabular}{ccc}
Coefficient & \multicolumn{2}{c}{Fit value} \\ 
&  $B_s\to D_s\ell \nu$ & $B\to 
D\ell \nu$ \\
\vspace*{-10pt}\\
\hline 
\vspace*{-8pt}\\
$a_0^{(0)}$ & 0.663(32) & 0.639(32)\\ 
$a_1^{(0)}$ & -0.10(30) & 0.18(33)\\ 
$a_2^{(0)}$ & 1.3(2.8) & -0.2(2.9) \\ 
$P_0$ & 6.43(10) & 6.43(10) \\
\\
$a_0^{(+)}$ & 0.868(34) & 0.870(38)\\ 
$a_1^{(+)}$ & -3.35(43) & -3.27(59)\\ 
$a_2^{(+)}$ & 0.6(4.7) & 0.5(4.8)\\
$P_+$ & 6.330(9) & 6.330(9) \\
\end{tabular}
\end{ruledtabular}
\end{table}

\begin{table}
\caption{\label{tab:Bszfit}
Group I priors and fit results for the parameters in the 
modified $z$-expansion for the 
$B_s\to D_s\ell \nu$ decay.
}
\begin{ruledtabular}
\begin{tabular}{ccccc}
& Prior $[f_0]$ & Fit result $[f_0]$ & Prior $[f_+]$ & Fit result $[f_+]$ 
\\
\vspace*{-10pt}\\
\hline 
\vspace*{-8pt}\\
$a_0$ & 0.0(3.0) & 0.663(32) & 0.0(5.0) & 0.868(34) \\ 
$a_1$ & 0.0(3.0) & -0.10(30) & 0.0(5.0) & -3.35(43) \\
$a_2$ & 0.0(3.0) & 1.3(2.8) & 0.0(5.0) & 0.6(4.7) \\ 
$c_1^{(1)}$ & 0.0(1.0) &  0.28(15) & 0.0(1.0) &  0.43(15) \\
$c_1^{(2)}$ & 0.0(1.0) & -0.20(1.0) & 0.0(1.0) & 0.48(62) \\ 
$c_1^{(3)}$ & 0.0(1.0) &  0.03(1.0) & 0.0(1.0) & -0.003(1.0) \\ 
$c_2^{(1)}$ & 0.00(30) &  0.20(13) & 0.00(30) & 0.31(13) \\
$c_2^{(2)}$ & 0.00(30) & 0.02(30) & 0.00(30) & -0.05(29) \\
$c_2^{(3)}$ & 0.00(30) & -0.005(0.3) & 0.00(30) & 0.0002(0.3) \\  
$d_1^{(1)}$ & 0.00(30) & -0.19(28) & 0.00(30) & -0.02(29) \\
$d_1^{(2)}$ & 0.00(30) & -0.003(0.3) & 0.00(30) & -0.002(0.3) \\
$d_1^{(3)}$ & 0.00(30) & 0.002(0.3) & 0.00(30) & -7$\times10^{-5}$(0.3) \\
$d_2^{(1)}$ & 0.00(30) & 0.04(30) & 0.00(30) & 0.05(30) \\
$d_2^{(2)}$ & 0.00(30) & -0.0002(0.3) & 0.00(30) & 0.003(0.3) \\
$d_2^{(3)}$ & 0.00(30) & 2$\times10^{-5}$(0.3) & 0.00(30) & 
-1$\times10^{-5}$(0.3) \\
$e_1^{(1)}$ & 0.00(30) & 0.22(24) & 0.00(30) & 0.08(24) \\
$e_1^{(2)}$ & 0.00(30) & -0.005(0.3) & 0.00(30) & -0.02(30) \\
$e_1^{(3)}$ & 0.00(30) & 0.004(0.3) & 0.00(30) & -0.0001(0.3) \\
$e_2^{(1)}$ & 0.0(1.0) & 1.42(53) & 0.0(1.0) & 0.70(73) \\
$e_2^{(2)}$ & 0.0(1.0) & -0.02(1.0) & 0.0(1.0) & -0.07(99) \\
$e_2^{(3)}$ & 0.0(1.0) & 0.009(1.0) & 0.0(1.0) & -0.0002(1.0) \\
$m_1^{(1)}$ & 0.00(30) & -0.007(0.236) & 0.00(30) & -0.05(22) \\
$m_1^{(2)}$ & 0.00(30) & -0.001(0.3) & 0.00(30) & -0.10(29) \\
$m_1^{(3)}$ & 0.00(30) & 0.009(0.3) & 0.00(30) & -0.0002(0.3) \\
$m_2^{(1)}$ & 0.0(1.0) & -0.43(42) & 0.0(1.0) & -0.17(38) \\
$m_2^{(2)}$ & 0.0(1.0) & 0.0003(1.0) & 0.0(1.0) & -0.77(85) \\
$m_2^{(3)}$ & 0.0(1.0) & 0.04(1.0) & 0.0(1.0) & -0.0004(1.0) \\
\end{tabular}
\end{ruledtabular}
\end{table}

\begin{table}
\caption{\label{tab:Bszfit2}
Group II priors and fit results for the parameters in the 
modified $z$-expansion for the 
$B_s\to D_s\ell \nu$ decay.
}
\begin{ruledtabular}
\begin{tabular}{ccc}
Quantity & Prior  & Fit result
\\
\vspace*{-10pt}\\
\hline 
\vspace*{-8pt}\\
$r_1/a$ & 2.6470(30) & 2.6474(30) \\
& 2.6180(30) & 2.6179(30) \\ 
& 2.6440(30) & 2.6437(30) \\
& 3.6990(30) & 3.6992(30) \\
& 3.7120(40) & 3.7116(39) \\
$aM_B$ & 3.23019(25) &  3.23018(25) \\
& 3.26785(33) &  3.26783(33) \\
& 3.23585(38) & 3.23579(38) \\
& 2.30884(17) & 2.30885(17) \\
& 2.30163(23) & 2.30162(22) \\
$aE_D(0,0,0)$ & 1.18750(15) &  1.18750(15) \\
& 1.20126(21) & 1.20125(20) \\
& 1.19031(24) & 1.19028(24) \\
&  0.84680(10) &  0.84680(10) \\
& 0.84410(12) & 0.84410(12) \\
$aE_D(1,0,0)$ & 1.21497(19) & 1.21506(19)  \\
& 1.24055(30) & 1.24075(28) \\
& 1.23055(35) &  1.23060(31)  \\
& 0.87579(16) & 0.87582(15) \\
& 0.87340(19) & 0.87338(19)  \\
$aE_D(1,1,0)$ & 1.24264(19) & 1.24276(19)  \\
& 1.27942(29) & 1.27953(27) \\
& 1.26974(35) &  1.26948(32)  \\
& 0.90397(16) & 0.90399(15) \\
& 0.90138(18) & 0.90135(18)  \\
$aE_D(1,1,1)$ & 1.26988(22) & 1.26999(22)  \\
& 1.31755(46) & 1.31737(40)  \\
&  1.30768(48) & 1.30738(41)   \\
& 0.93131(21) & 0.93132(20) \\
& 0.92861(24) & 0.92864(23) \\
$aM_\pi$ &  0.15990(20) &  0.15990(20)    \\
& 0.21110(20) & 0.21110(20)  \\
& 0.29310(20) & 0.29310(20)  \\
&  0.13460(10) & 0.13460(10)  \\
& 0.18730(10) & 0.18730(10) \\
$aM_{\eta_s}$ & 0.41113(18) &  0.41113(18)  \\
&  0.41435(22) & 0.41435(22)  \\
&  0.41185(22) & 0.41185(22) \\
& 0.29416(12) &  0.29416(12)  \\
&  0.29311(18) & 0.29311(18)  \\
$aM_K$ &   0.31217(20) & 0.31217(20)   \\
& 0.32851(48) & 0.32850(48)  \\
& 0.35720(22) & 0.35721(22)  \\
& 0.22855(17) & 0.22855(17)  \\
& 0.24596(14) & 0.24596(14)   \\
$aM_K^{\mathrm{MILC}}$ &  0.36530(29) & 0.36530(29)  \\
& 0.38331(24) & 0.38331(24)  \\
& 0.40984(21) & 0.40984(21)  \\
& 0.25318(19) &  0.25318(19)  \\
& 0.27217(21) & 0.27217(21)  \\
$aM_\pi^{\mathrm{MILC}}$ &  0.15971(20) &  0.15971(20)  \\
& 0.22447(17) & 0.22447(17)  \\
& 0.31125(16) & 0.31125(16)  \\
& 0.14789(18) & 0.14789(18)   \\
& 0.20635(18) &  0.20635(18)   \\
$1+m_\parallel$    & 1.000(30) &    1.001(30)   \\
$1+m_\perp$ &  1.000(30) &       1.000(30)
\end{tabular}
\end{ruledtabular}
\end{table}

\begin{table}
\caption{\label{tab:Bszfit3}
Group III priors and fit results for the parameters in the 
modified $z$-expansion for the 
$B_s\to D_s\ell \nu$ decay.
}
\begin{ruledtabular}
\begin{tabular}{ccc}
Quantity & Prior (GeV)  & Fit result (GeV)
\\
\vspace*{-10pt}\\
\hline 
\vspace*{-8pt}\\
$r_1$ & 0.3133(23) & 0.3130(23) \\
$m_{\eta_s}^{\mathrm{phys}}$ & 0.6858(40) & 0.6858(40) \\
$m_\pi^{\mathrm{phys}}$ & 0.13500000(60) & 0.13500000(60) \\
$m_{B_s}^{\mathrm{phys}}$ & 5.36679(23) & 5.36679(23) \\
$m_{D_s}^{\mathrm{phys}}$ & 1.96830(10) & 1.96830(10) \\
$m_{K_s}^{\mathrm{phys}}$ & 0.4957(20) & 0.4957(20) \\
$M_+$ & 6.3300(90) & 6.3300(90) \\
$M_0$ & 6.398(99) & 6.42(10) \\
\end{tabular}
\end{ruledtabular}
\end{table}

\begin{table*}
\caption{\label{tab:BBszfit}
Group I priors and fit results for the parameters in the 
modified $z$-expansion for the ratio of the form factors for the 
$B_s\to D_s\ell \nu$ decay, indicated by the superscript $B_s$, and $B\to D\ell 
\nu$ decay, labeled by the superscript $B$.
}
\begin{ruledtabular}
\begin{tabular}{ccccccccc}
& Prior $[f_0^{B_s}]$ & Fit result $[f_0^{B_s}]$ & Prior 
$[f_+^{B_s}]$ & Fit result $[f_+^{B_s}]$
& Prior $[f_0^{B}]$ & Fit result $[f_0^{B}]$ & Prior 
$[f_+^{B}]$ & Fit result $[f_+^{B}]$
\\
\vspace*{-10pt}\\
\hline 
\vspace*{-8pt}\\
$a_0$ & 0.0(3.0) & 0.663(32) & 0.0(5.0) & 0.639(32) 
& 0.0(3.0) & 0.868(34) & 0.0(5.0) & 0.870(38) \\ 
$a_1$ & 0.0(3.0) & -0.10(30) & 0.0(5.0) & 0.18(33) 
& 0.0(3.0) & -3.35(43) & 0.0(5.0) & -3.27(59)\\
$a_2$ & 0.0(3.0) & 1.3(2.8) & 0.0(5.0) & -0.2(2.9) 
& 0.0(3.0) & 0.6(4.7) & 0.0(5.0) & 0.5(4.8) \\ 
$c_1^{(1)}$ & 0.0(1.0) & 0.28(15) & 0.0(1.0) &  -0.10(23) 
& 0.0(1.0) &  0.43(15) & 0.0(1.0) &  0.50(25) \\
$c_1^{(2)}$ & 0.0(1.0) & -0.2(1.0) & 0.0(1.0) & -0.08(1.0) 
& 0.0(1.0) & 0.48(62) & 0.0(1.0) & -1.13(79) \\ 
$c_1^{(3)}$ & 0.0(1.0) &  0.03(1.0) & 0.0(1.0) & 0.002(1.0) 
 & 0.0(1.0) &  -0.003(1.0) & 0.0(1.0) & 0.004(1.0) \\ 
$c_2^{(1)}$ & 0.00(30) &  0.20(13) & 0.00(30) & -0.11(19) 
& 0.00(30) &  0.31(13) & 0.00(30) & 0.38(20) \\
$c_2^{(2)}$ & 0.00(30) & 0.02(30) & 0.00(30) & 0.008(0.3) 
& 0.00(30) & -0.05(29) & 0.00(30) & 0.13(29) \\
$c_2^{(3)}$ & 0.00(30) & -0.005(0.3) & 0.00(30) & -0.0003(0.3) 
& 0.00(30) & 0.0002(0.3) & 0.00(30) & -0.0005(0.3) \\  
$d_1^{(1)}$ & 0.00(30) & -0.19(28) & 0.00(30) & 0.01(28) 
& 0.00(30) & -0.02(29) & 0.00(30) & -0.06(28) \\
$d_1^{(2)}$ & 0.00(30) & -0.003(0.3) & 0.00(30) & 0.0005(0.3) 
& 0.00(30) & -0.002(0.299) & 0.00(30) & -0.02(0.3) \\
$d_1^{(3)}$ & 0.00(30) & 0.002(0.3) & 0.00(30) & 2$\times10^{-5}$(0.3) 
& 0.00(30) & -7$\times10^{-5}$(0.3) & 0.00(30) & 9$\times10^{-5}$(0.3) \\
$d_2^{(1)}$ & 0.00(30) & 0.04(30) & 0.00(30) & -0.02(30) 
 & 0.00(30) & 0.05(30) & 0.00(30) & 0.06(30) \\
$d_2^{(2)}$ & 0.00(30) & -0.0002(0.3) & 0.00(30) & -0.0003(0.3) 
& 0.00(30) & 0.003(0.3) & 0.00(30) & -0.002(0.3) \\
$d_2^{(3)}$ & 0.00(30) & 2$\times10^{-5}$(0.3) & 0.00(30) & 
3$\times10^{-6}$(0.3) 
 & 0.00(30) & 2$\times10^{-5}$(0.3) & 0.00(30) & 
-1$\times10^{-6}$(0.3)\\
$e_1^{(1)}$ & 0.00(30) & 0.22(24) & 0.00(30) & 0.27(25) 
& 0.00(30) & 0.08(24) & 0.00(30) & 0.05(25) \\
$e_1^{(2)}$ & 0.00(30) & -0.005(0.3) & 0.00(30) & 0.006(0.3)
& 0.00(30) & -0.02(0.3) & 0.00(30) & -0.01(30) \\
$e_1^{(3)}$ & 0.00(30) & 0.004(0.3) & 0.00(30) & -$8\times10^{-5}$(0.3) 
& 0.00(30) & -0.0001(0.3) & 0.00(30) & $4\times10^{-5}$(0.3)\\
$e_2^{(1)}$ & 0.0(1.0) & 1.42(53) & 0.0(1.0) & 1.49(66)
& 0.0(1.0) & 0.70(73) & 0.0(1.0) & 0.12(82) \\
$e_2^{(2)}$ & 0.0(1.0) & -0.02(1.0) & 0.0(1.0) & 0.02(1.0)
& 0.0(1.0) & -0.07(1.0) & 0.0(1.0) & -0.02(99) \\
$e_2^{(3)}$ & 0.0(1.0) & 0.009(1.0) & 0.0(1.0) & -0.0003(1.0)
 & 0.0(1.0) & -0.0002(1.0) & 0.0(1.0) & $3\times10^{-5}$(1.0) \\
$m_1^{(1)}$ & 0.00(30) & -0.007(0.236) & 0.00(30) & -0.10(24)
& 0.00(30) & -0.05(22) & 0.00(30) & 0.03(24) \\
$m_1^{(2)}$ & 0.00(30) & -0.001(0.3) & 0.00(30) & 0.02(30)
 & 0.00(30) & -0.10(29) & 0.00(30) & -0.03(29) \\
$m_1^{(3)}$ & 0.00(30) & 0.009(0.3) & 0.00(30) & -0.0003(0.3)
& 0.00(30) & -0.0002(0.3) & 0.00(30) & $5\times10^{-5}$(0.3)\\
$m_2^{(1)}$ & 0.0(1.0) & -0.43(42) & 0.0(1.0) & -0.31(44) 
 & 0.0(1.0) & -0.17(38) & 0.0(1.0) & -0.19(40)\\
$m_2^{(2)}$ & 0.0(1.0) & 0.0003(1.0) & 0.0(1.0) & 0.1(1.0) 
& 0.0(1.0) & -0.77(85) & 0.0(1.0) & -0.12(89) \\
$m_2^{(3)}$ & 0.0(1.0) & 0.04(1.0) & 0.0(1.0) & -0.002(1.0) 
& 0.0(1.0) & -0.0004(1.0) & 0.0(1.0) & $5\times10^{-5}$(1.0) \\
\end{tabular}
\end{ruledtabular}
\end{table*}

\begin{table*}
\caption{\label{tab:BBszfit2}
Group II priors and fit results for the parameters in the 
modified $z$-expansion for the ratio of the form factors for the 
$B_s\to D_s\ell \nu$ and $B\to D\ell \nu$ decays.
}
\begin{ruledtabular}
\begin{tabular}{ccccc}
Quantity & Prior $[B_s\to D_s\ell \nu]$  & Fit result $[B_s\to D_s\ell \nu]$   
& Prior $[B\to D\ell \nu]$   & Fit result $[B\to D\ell \nu]$  \\
\vspace*{-10pt}\\
\hline 
\vspace*{-8pt}\\
$aM_{B_{(s)}}$ & 3.23019(25) & 3.23017(25)  & 3.18937(62) & 3.18933(62) \\
& 3.26781(33) & 3.26782(33)& 3.23194(88) & 3.23211(87)\\
& 3.23575(38) & 3.23578(38) & 3.21199(77) & 3.21193(77) \\
& 2.30906(26) & 2.30905(26) & 2.28120(49) & 2.28117(48)  \\
& 2.30122(16) & 2.30122(16) & 2.28102(40) & 2.28112(40) \\
$aE_{D_{(s)}}(0,0,0)$ & 1.18750(15) &  1.18750(15) &  1.13904(97) & 1.13927(84)  
\\ 
&  1.20126(21) & 1.20126(20)  &  1.16001(73) &   1.16026(71) \\ 
& 1.19031(24) & 1.19026(24) &  1.16339(54) &    1.16333(54)      \\ 
& 0.84675(12) & 0.84674(10)  &  0.81448(35)&    0.81444(35)     \\ 
&  0.84419(10) & 0.84421(10) &  0.81995(27)   &   0.82005(26)    \\ 
$aE_{D_{(s)}}(1,0,0)$ & 1.21497(19) & 1.21505(19)  &  1.1682(10) &   1.16794(90) 
  \\ 
&  1.24055(30) & 1.24076(28) & 1.19896(99) &     1.19915(94)     \\ 
&  1.23055(35) & 1.23058(31) & 1.20399(76)  &    1.20448(69)   \\ 
&  0.87579(16) & 0.87580(15) & 0.84377(56)&      0.84399(50) \\ 
&  0.87353(16) & 0.87344(15) & 0.85102(40) &  0.85086(38)      \\ 
$aE_{D_{(s)}}(1,1,0)$ & 1.24264(19) & 1.24275(19) & 1.19863(85)   &   
1.19853(82)    \\ 
&  1.27942(29) & 1.27953(27)  & 1.24009(87)    &    1.23987(83)     \\ 
& 1.26974(35) & 1.26945(32)& 1.24476(78)& 1.24471(72)   \\ 
&  0.90397(16) & 0.90398(15)   & 0.87274(56) & 0.87267(52)  \\  
&  0.90144(16) & 0.90146(15)& 0.87943(38)  & 0.87950(36)    \\ 
$aE_{D_{(s)}}(1,1,1)$ &  1.26988(22) & 1.26998(22)  & 1.22850(85)  &   
1.22833(83)    \\  
&  1.31755(46) & 1.31732(40) & 1.27838(93) & 1.27815(91)    \\ 
& 1.30768(48) & 1.30751(42)   & 1.28312(97) & 1.28316(90)      \\ 
& 0.93126(24) & 0.93126(24)  & 0.89996(74) & 0.90037(66)    \\  
&  0.92873(24) & 0.92879(20) & 0.90647(50)& 0.90645(47)      \\ 
\end{tabular}
\end{ruledtabular}
\end{table*}

\begin{table}
\caption{\label{tab:BBszfit3}
Shared (Group II and III) priors and fit results for the parameters in the 
modified $z$-expansion for the ratio of the form factors for the 
$B_s\to D_s\ell \nu$ and $B\to D\ell \nu$ decays. These priors are common to 
both fits to the $B_s\to D_s\ell \nu$ and $B\to D\ell \nu$ decays, which are 
fitted in the same script to account for correlations between form factor 
results. Values for Group III priors are given in GeV.
}
\begin{ruledtabular}
\begin{tabular}{ccc}
Quantity & Prior  & Fit result
\\
\vspace*{-10pt}\\
\hline 
\vspace*{-8pt}\\
$r_1/a$ & 2.6470(30) & 2.6474(30) \\
& 2.6180(30) & 2.6174(30) \\
& 2.6440(30) & 2.6442(30) \\
& 3.6990(30) & 3.6990(30)  \\
& 3.7120(40) & 3.7121(39) \\
$1+m_\parallel$    & 1.000(30) &    0.998(30)   \\
$1+m_\perp$ &  1.000(30) &       1.003(30)\\
\vspace*{-10pt}\\
\hline 
\vspace*{-8pt}\\
Quantity & Prior (GeV) & Fit result (GeV)
\\
\vspace*{-10pt}\\
\hline 
\vspace*{-8pt}\\
$r_1$ & 0.3132(23) & 
0.3130(23) \\
$m_{\eta_s}^{\mathrm{phys}}$ & 0.6858(40) & 0.6858(40) \\
$m_\pi^{\mathrm{phys}}$ & 0.13500000(60) & 0.13500000(60) \\
$m_{B_s}^{\mathrm{phys}}$ & 5.36679(23) & 5.36679(23) \\
$m_{D_s}^{\mathrm{phys}}$ & 1.96830(10) & 1.96830(10) \\
$m_{K_s}^{\mathrm{phys}}$ & 0.4957(20) & 0.4957(20) \\
$m_B^{\mathrm{phys}}$ & 5.27941(17) & 5.27942(17) \\
$m_D^{\mathrm{phys}}$ & 1.86690(40) & 1.86690(40) \\
$M_+$ & 6.3300(90) & 6.3300(90) \\
$M_0$ & 6.42(10) & 6.42(10) \\
\end{tabular}
\end{ruledtabular}
\end{table}

%-----------------------------------o
%  References
%-----------------------------------o
\clearpage
%\bibliography{mbw}
\bibliographystyle{elsart-num}
\bibliographystyle{apsrev}
\bibliography{bstods.bib}

%======================================================= The End ========8

\end{document}